\documentclass[
reprint,
amsmath,amssymb,
aip,
jcp,
floats,
]{revtex4-1}
\usepackage{xcolor}
\usepackage{color}
\usepackage{amssymb, amsmath}
\usepackage{bm}
\usepackage{graphicx}
\usepackage[caption=false]{subfig}
\usepackage{fancyhdr}
\usepackage{float}

\begin{document}
\rhead{\today}
\cfoot{\thepage}
\title{Accurate Non-adiabatic Quantum Dynamics from Pseudospectral Sampling of Time-dependent Gaussian Basis Sets}
	\author{Charles W. Heaps and David A. Mazziotti}
	\email{damazz@uchicago.edu}
	\affiliation{Department of Chemistry and The James Franck Institute, The University of Chicago, Chicago, Illinois, 60637, United States}
	\date{Submitted March 25, 2016; Revised June 15, 2016}
	\begin{abstract}
	Quantum molecular dynamics requires an accurate representation of the molecular potential energy surface from a minimal number of electronic structure calculations, particularly for nonadiabatic dynamics where excited states are required.   In this paper, we employ pseudospectral sampling of time-dependent Gaussian basis functions for the simulation of non-adiabatic dynamics.  Unlike other methods, the pseudospectral Gaussian molecular dynamics tests the Schr\"{o}dinger equation with $N$ Dirac delta functions located at the centers of the Gaussian functions reducing the scaling of potential energy evaluations from $\mathcal{O}(N^2)$ to $\mathcal{O}(N)$.  By projecting the Gaussian basis onto discrete points in space, the method is capable of efficiently and quantitatively describing nonadiabatic population transfer and intra-surface quantum coherence. We investigate three model systems; the photodissociation of three coupled Morse oscillators, the bound state dynamics of two coupled Morse oscillators, and a two-dimensional model for collinear triatomic vibrational dynamics.  In all cases, the pseudospectral Gaussian method is in quantitative agreement with numerically exact calculations.  The results are promising for nonadiabatic molecular dynamics in molecular systems where strongly correlated ground or excited states require expensive electronic structure calculations.
	\end{abstract}
\maketitle

\section{Introduction}
	For the vast majority of molecular dynamics, the Born-Oppenheimer approximation is valid and the nuclear evolution may be described by a single electronic potential energy surface.  However, nonadiabatic dynamics, or cases when the Born-Oppenheimer approximation breaks down, characterize many important reactions in chemistry, from charge transfer in materials to photo-induced biological processes.\cite{doi:10.1021/jp962134y, doi:10.1146/annurev.physchem.49.1.125, tully:22A301, persico:2014, doi:10.1021/ar500357y}  Unfortunately, the \emph{ab initio} description of nonadiabatic chemical processes is still a significant challenge for computation because of the accuracy and efficiency required for both the quantum molecular dynamics and the molecular electronic structure.\cite{BenNun199857,BenNun:1999a,doi:10.1021/jp994174i,ben-nun:2002,:/content/aip/journal/jcp/122/11/10.1063/1.1871876,Lasorne2006604,B700297A,:/content/aip/journal/jcp/133/18/10.1063/1.3504167,tully:22A301,:/content/aip/journal/jcp/137/22/10.1063/1.4734313,leveque:044320,Makhov:2014, persico:2014,lasorne:2014} Electronic structure calculations for nonadiabatic dynamics are computationally expensive because they must generate accurate excited electronic states with a balanced description of strong correlation.\cite{doi:10.1021/ar960091y,doi:10.1021/jp072027b,:/content/aip/journal/jcp/131/12/10.1063/1.3237029,doi:10.1021/jp9063565,:/content/aip/journal/jcp/132/2/10.1063/1.3275806,Mazziotti2012}  A time-dependent (trajectory-guided) basis set circumvents the exponential scaling of traditional grid methods while being compatible with the direct determination of the potential energy surface from electronic structure calculations.   In this work, we implement the nonadiabatic extension of a recently introduced trajectory-guided Gaussian basis set for quantum molecular dynamics called pseudospectral Gaussian dynamics.~\textcolor{black}{\cite{Heaps:2016}}  The most important advantage of pseudospectral Gaussian dynamics is the ability to match the accuracy of analytical potential energy integration with only $\mathcal{O}(N)$ sampling where $N$ is the number of basis functions of the potential energy surface.
	
A particularly important component of a nonadiabatic trajectory-guided basis set method is an efficient approach to capturing the coherence and decoherence between trajectories on different potential energy surfaces.  To achieve this goal in the present work, we employ an efficient representation of the potential energy surfaces in the time-dependent Schr{\"o}dinger equation that uses pseudospectral sampling with time-dependent Gaussian basis functions.   In traditional Gaussian-based \textcolor{black}{spectral} methods, the basis set is tested using the complex conjugate of the basis functions,\cite{Huber:1988,doi:10.1021/jp953105a,Shalashilin:2000,Mauritz:2001,Wu:2003,PhysRevLett.110.263202,doi:10.1021/ct500657f} \textcolor{black}{which requires integral evaluation over all space, a nontrivial task for the potential energy surface}.  Generally, approximations \textcolor{black}{to the potential energy surface} such as the local harmonic approximation (LHA), saddle point approximation (SPA) and the bra-ket averaged Taylor expansion (BAT) are introduced in the Hamiltonian,\cite{Heller:1975,doi:10.1021/jp953105a,A605958I,doi:10.1021/jp970842t,BenNun199857,Worth2003502,B700297A,:/content/aip/journal/jcp/137/22/10.1063/1.4734313,Makhov:2014, C5CP07332D}  but they rely on the locality of the Gaussian basis functions to ensure accuracy and often require additional electronic structure calculations beyond the number of Gaussian basis-set functions.  \textcolor{black}{While the spectral Gaussian dynamics uses the same Gaussian functions for the basis functions (functions in the expansion of the ket) and the test functions (functions in the expansion of the bra), the pseudospectral Gaussian dynamics employs Dirac delta functions as the test functions}.\cite{boyd2001chebyshev, Lill1982483,Kosloff198335,Yang198898, doi:10.1021/j100319a003, Peet:1989a, Peet:1989, Peet:1990a, B814315C}  The delta function test function reduces the integral evaluation to function evaluation.\cite{Yang198898,Peet:1989a,Peet:1989,Yang:1989b,Peet:1990a,Dehghan2007136,B814315C}  The potential energy surfaces can then be expressed accurately with $\mathcal{O}(N)$ scaling, meaning that the electronic structure information for the basis function trajectories is required to incorporate the quantum dynamics.  \textcolor{black}{The use of Dirac delta test functions in conjunction with a spectral basis-set expansion is a key feature of the pseudospectral method.}\cite{orszag:1969,gottlieb1977numerical, boyd2001chebyshev, furnaro:1992,fornberg1998practical,Canuto:2006,hesthaven2007spectral, tannor2007introduction}

	Having introduced a prescription to represent the Hamiltonian efficiently, we need to prescribe the time dependence of the basis functions.  While classical equations of motion are often sufficient for single-surface dynamics, nonadiabatic population transfer precludes the straightforward definition of a single, classical force.  In this work, we use Ehrenfest trajectories\cite{BILLING1983535, tully:1061, A801824C, Hack:2000,
:/content/aip/journal/jcp/130/24/10.1063/1.3153302, Subotnik:2010fk} to propagate the basis functions.  While Ehrenfest trajectories are suitable for some of the applications we present, it is well-known that they provide a qualitatively incorrect description of nonadiabatic processes when the gradients of the multiple potential energy surfaces differ significantly.  The consequence is that many more trajectories are required for convergence than would likely be needed if an improved selection of trajectories was made.  The {\em ab initio} multiple spawning (AIMS) algorithm\cite{Martinez:1996,A605958I,ben-nun:2002,doi:10.1021/ar040202q,:/content/aip/journal/jcp/130/13/10.1063/1.3103930} and the fewest switches surface hopping method\cite{tully:1061, A801824C, Subotnik:2011fk} are very effective approaches to minimizing basis set size that would be expected to accelerate convergence.  Nevertheless, we observe well-behaved convergence to the exact solution for one- and two-dimensional examples using Ehrenfest trajectories.
	
	The discretized grid of nonadiabatic pseudospectral Gaussian dynamics places the method at the intersection of independent trajectory, Gaussian basis set expansion, and nonadiabatic Bohmian methods.  While we adopt a basis of independently propagated Gaussians as in AIMS, the numerical framework resembles that of a Bohmian grid where each trajectory corresponds to a discrete point in space with an associated amplitude of the wavefunction.\cite{wyatt:5113,PhysRevA.71.032511,C0CP02175J,zamstein:22A517,zamstein:22A518,curchod:184112}    By associating a Gaussian basis function with each Bohmian trajectory, we are able to build a matrix form of the time-dependent Schr\"{o}dinger equation that circumvents the challenging spacial derivatives and  nodal instabilities in Bohmian mechanics.\cite{wyatt:5113,PhysRevA.71.032511,C0CP02175J,zamstein:22A517,zamstein:22A518,curchod:184112}

Many of the methods to describe decoherence in fewest-switches surface hopping (FSSH) also invoke a Gaussian form for individual trajectories.\cite{Bittner:1995,:/content/aip/journal/jcp/133/13/10.1063/1.3489004,Subotnik:2010fk,Subotnik:2011fk,landry:191101,doi:10.1021/jp206557h,:/content/aip/journal/jcp/135/2/10.1063/1.3603447,Shenvi:2011uq,Subotnik:2011uq,landry:22A513}  In FSSH the Gaussian overlap allows the straightforward calculation of overlap decoherence criteria and a route to an improved description of quantum-mechanical behavior.   Since the pseudospectral Gaussian method, without approximation, requires only the information used in FSSH propagation, the construction of the full Hamiltonian may be viewed as a means of coupling all trajectories simultaneously.  The main computational cost is the requirement to solve a system of equations $Ax=b$ at each time step.  However, if solving the linear system of equations provides converged results using many fewer trajectories, the calculation time may be dominated by electronic structure calculations in on-the-fly applications rather than the solution of the linear system of equations.
	
	 The discretization here should not be confused with the phase space discretization discussed in some coherent state methods.\cite{Shalashilin:2000,Mauritz:2001, Wu:2003}  In those cases, position and momenta are discretized to approximate the complete set of coherent states, but the Schr{\"o}dinger equation is still tested with the basis functions requiring the approximation of potential integrals.  In this paper, we also begin with a discretized basis set in phase space, but we then project the Schr{\"o}dinger equation onto discrete points in space.  Although the pseudospectral approach has been previously employed in chemical dynamics in the context of non-Gaussian dynamics, trajectory-guided Gaussian methods have been predominantly solved in the functional space of the Gaussian basis sets rather than a space of Dirac delta functions.
	
In the present paper, we build upon our recent work of applying the pseudospectral Gaussian method to a single Born-Oppenheimer potential energy surface by studying nonadiabatic model systems in the diabatic representation.\cite{Heaps:2016}  Previously, we demonstrated the effectiveness of the pseudospectral Gaussian method for adiabatic dynamics in as many as six dimensions.  Although we study one- and two-dimensional nonadiabatic systems in this work, the previous results are promising for the future application of the method to high-dimensional systems.

In previous work, we also investigated the performance of pseudospectral sampling to the BAT approximation, an $\mathcal{O}(N)$ approximation implemented in recent Gaussian-based methods.\cite{:/content/aip/journal/jcp/137/22/10.1063/1.4734313,Makhov:2014, C5CP07332D}  While the BAT does reduce the scaling from original Gaussian methods, it is still an approximation to a potential energy integral over all space.  We demonstrate that although the BAT provides accurate dynamics in some cases, it breaks down when non-local effects become particularly important.  The breakdown arises from the series expansion to the integral and the implicit requirement for localized basis functions and dynamics.
	
	  We review the equations of the pseudospectral Gaussian method including the Hamiltonian elements required to couple surfaces and equations of motion for the basis functions.  The first two applications use models of coupled Morse potentials.  We consider a set of three-surface photodissociation models and a two-surface case of bound excitation leading to anharmonic oscillation.\cite{Coronado2001521, Bonella:2003}   The third case studied is a two-dimensional model corresponding to vibrational dynamics in a collinear triatomic molecule.\cite{Ferretti:1996,:/content/aip/journal/jcp/130/13/10.1063/1.3103930}      Collectively, these model systems allow us to test many important features including multiple crossings through regions of coupling, spatial separation of surface densities, and intra-surface coupling effects.  Not only is the pseudospectral Gaussian method able to describe population dynamics accurately, it also allows for accurate wave packet reconstruction after long propagation times.

\section{Theory}
\label{sec:theory}
The pseudospectral Gaussian method is reviewed with a discussion of the test functions, basis functions and associated equations of motion in Section \ref{subsec:test_functions}.  In Section \ref{subsec:matrix_eq} we present the matrix form of the working equations is for nonadiabatic dynamics in the diabatic representation.
\subsection{Test and Basis Functions}
\label{subsec:test_functions}
Testing the time-dependent Schr{\"o}dinger equation with the $N_f$ test functions $\chi_{i}$ yields
		\begin{equation}
			\left\langle \chi_i(\mathbf{r,x},t) \left| i\frac{d}{d t} - \hat H \right| \Psi(\mathbf{r,x},t) \right\rangle = 0.
			\label{eq:tdse}
		\end{equation}

We may expand the total wavefunction using the Born-Huang expansion in the basis of orthonormal electronic states,\cite{born1998dynamical}
	\begin{equation}
		\Psi(\mathbf{r,x},t)=\sum_{I=1}^{\infty} C^I(t)\Omega^I(\mathbf{r,x})\Phi^I(\mathbf{x},t),
	\end{equation}
where $\Omega^{I}(\mathbf{r,x})$ is the ${I}^{\rm th}$ electronic wavefunction that depends parametrically on nuclear coordinates ${\bf x}$ and $\Phi^I(\mathbf{x},t)$ is the time-dependent nuclear wavefunction of the ${I}^{\rm th}$  electronic state.
	Throughout the paper ${\bf r}$ denotes the fast (electronic) coordinates and ${\bf x}$ denotes the slow (nuclear) coordinates.
	Superscripts and capital letters denote electronic states while subscripts and lower case letters denote primitive Gaussians in the nuclear wavefunction expansion for a given electronic state.
	
The wavefunction for the $I$th electronic state can be expanded in terms of $N_f$ basis functions $\phi_{j}$,
	\begin{equation}
		 \Phi^I(\mathbf{x},t) = \sum_j^{N_f} c_j(t) \phi_j(\mathbf{x},t).
		\label{eq:basis_exp}
	\end{equation}
Since the wave function is a finite approximation, there will be a nonzero residual between the approximate and exact solution.  One may prescribe a method to minimize the norm of the residual through the choice of the functions upon which the partial differential equation is projected, known as the test functions.\cite{finlayson1972method, boyd2001chebyshev}  If the set of basis functions and the set of test functions are chosen to be the same (in this case, Gaussian functions), known as a spectral method, then the evaluation of the potential requires numerical integration which, without approximation, scales as $\mathcal{O}(N_f^2)$.  However, if we choose the test functions to be Dirac delta functions, located at the centers of the Gaussian basis functions, then the evaluation of the potential scales as $\mathcal{O}(N_f)$.  Testing the basis set expansion with Dirac delta functions is known as the pseudospectral method.   Because the representation of the Hilbert space retains the use of the Gaussian basis functions with the test functions placed at the centers of these functions, the approximation is a pseudospectral Gaussian method.  \textcolor{black}{While using a set of test functions that is distinct from the set of basis-set functions generates a non-Hermitian Hamiltonian matrix, a Hermitian Hamiltonian operator can be accurately represented by a non-Hermitian Hamiltonian matrix, as first noted in a series of theoretical chemistry papers by Frost~\cite{F64}.  Gottlieb and coworkers~\cite{hesthaven2007spectral} have demonstrated the accuracy of pseudospectral methods for general time-dependent problems on unstructured grids.}  In Section~\ref{subsec:matrix_eq}, we formulate a pseudospectral version of the common trajectory-guided Gaussian-basis Hamiltonian.  The Gaussian functions are chosen to be time-dependent, moving according to Hamilton's equations of motion.   For clarity, we introduce $D^I_{j}(t) = C^I(t) c_{j}(t)$ where both the electronic-state amplitude and single-state expansion coefficient have been absorbed into a single expansion coefficient.  \textcolor{black}{As demonstrated in Ref.~\onlinecite{Heaps:2016}, the pseudospectral method for Gaussian dynamics is as accurate as the traditional spectral method.}

Each basis function, for a problem in $N_d$-dimensions, is given as a product of one-dimensional functions
		\begin{equation}
		\begin{aligned}
			\label{eq:basis_func}
			&\phi_j(\mathbf{x},t; \mathbf{\alpha}_j, \mathbf{x}_j(t),\mathbf{p}_j(t), \gamma_j(t)) = \\
			&\exp(\gamma_j) \prod_{k=1}^{N_{d}} \mathrm{exp}(-\alpha_{k_j}(\Delta x_{k_j})^2 + i p_{k_j}(\Delta x_{k_j})),
		\end{aligned}
	\end{equation}
	where $\Delta x_{k_j} = (x_k - x_{k_j})$ and \textcolor{black}{the width $\alpha_j$ is time-independent, which is known as the frozen Gaussian approximation.}\cite{Heller:1981}  The parameter $\gamma_j$ is complex, accounting for phase and normalization and determined by the local harmonic approximation.\cite{Heller:1975}  $N_f$ is the basis set size while $N_d$ is the number of degrees of freedom in the system. Therefore, $\mathbf{x}_j$ and $\mathbf{p}_j$ represent the $N_d$-dimensional vectors corresponding to the time-dependent basis function position and momentum centers for the $j$th basis function.

 The equations of motion for the $j$th basis function are given by
	\begin{subequations}
		\begin{align}
			\frac{\partial x_{k_j}}{\partial t} &= \frac{p_{k_j}}{m_k} \label{eq:subeom1} \\
			\frac{\partial p_{k_j}}{\partial t} &=  -\left.\frac{\partial V_{\mathrm{Ehr}}(\mathbf{x})}{\partial x_k}\right|_{x_{k_j}}  \\
			\frac{\partial \gamma_{j}}{\partial t} &= -i\left(V^{\mathrm{Ehr}}(\mathbf{x}_j)+\sum_k^{N_d}[2\alpha_{k_j} - p_{k_j}^2]/2m_k \right).\label{eq:subeom3}
		\end{align}
		\label{eq:basis_eom}
	\end{subequations}
The trajectories are determined by the Ehrenfest potential energy, defined by the state averaged Hamiltonian,
		\begin{equation}
		\begin{aligned}
			&V^{\mathrm{Ehr}}(\mathbf{x}_j) = \\
			&\frac{|D^1_j|^2 V_1(\mathbf{x}_j) + |D^2_j|^2 V_2(\mathbf{x}_j) + 2\mathrm{Re}(D^{1*}_j D^2_j V_{12}(\mathbf{x}_j))}{|D^1_j|^2 + |D^2_j|^2},
		\end{aligned}
		\end{equation}
which has been written explicitly for basis function $j$ in a two-level system.\cite{BILLING1983535, :/content/aip/journal/jcp/130/24/10.1063/1.3153302}
	
\subsection{Matrix Equations and their Solution}
\label{subsec:matrix_eq}
Assigning the $N_d$-dimensional Dirac delta function to the test function
	\begin{equation}
		 \chi_{i}^{I}(\mathbf{r,x},t)  = \delta(\mathbf{x} - \mathbf{x}_i) \Omega^I(\mathbf{r,x})
	\end{equation}
allows us to recast Eq.~(\ref{eq:tdse}) as the following matrix equation
	\begin{equation}
		\mathbf{\dot{D}}^I = -i \mathbf{\Phi}^{-1}\left(\mathbf{H}^{II} - i\mathbf{\dot{\Phi}}\right)\mathbf{D}^I -i \sum_{J, I\neq J} \mathbf{H}^{IJ} \mathbf{D}^J,
		\label{eq:matrix_eq}
	\end{equation}
	where the first term on the RHS of Eq.~(\ref{eq:matrix_eq}) accounts for intra surface coupling and the second term, $I\neq J$, accounts for inter surface coupling.
	The corresponding matrix elements are
		\begin{equation}
		\Phi_{ij} = \phi_j(\mathbf{x}_i)
	\end{equation}
	\begin{equation}
		{\dot \Phi_{ij}} = \sum_{k=1}^{N_d} \left. \frac{d \phi_j}{dt}\right|_{x_{k_i}}
	\end{equation}
	\begin{equation}
		H_{ij}^{II}  = -\sum_{k=1}^{N_d} \left.\frac{1}{2m_k}  \frac{\partial^2\phi_j}{\partial x_k^2}\right|_{x_{k_i}} + V^{II}(\mathbf{x}_i)\phi_j(\mathbf{x}_i) .
	\end{equation}
	\begin{equation}
		H_{ij}^{IJ}  = V^{IJ}(\mathbf{x}_i)\phi_j(\mathbf{x}_i) .
	\end{equation}
	The matrix  $\Phi$ is a discrete version of the overlap matrix.  The spatial and time derivatives of $\phi_i$ are simply calculated by taking the appropriate derivatives of Eq.~(\ref{eq:basis_func}).  Since all calculations are run in the diabatic representation in this paper, the derivative coupling terms are omitted for clarity.  All coupling between the surfaces occurs through the off-diagonal elements of the potential energy, $V^{IJ}$.\cite{A801824C,ben-nun:2002,tannor2007introduction,curchod:184112} Although all of the models studied here employ the diabatic representation, the main theoretical results are readily extended to the adiabatic representation, which will be pursued in future on-the-fly work.

Although Eq.~(\ref{eq:matrix_eq}) may be ill-conditioned, it can be readily solved for an accurate set of expansion coefficients $\{D^I_{j}\}$ through regularization methods for inverse problems.\cite{hansen2010discrete} We employ a singular value decomposition (SVD) with a threshold for removing small singular values.  Similar regularization methods are employed for the Gaussian-based methods with Gaussian test functions.\cite{ben-nun:7244, Burghardt:2003, PhysRevLett.110.263202, doi:10.1021/ct500657f}

\section{Applications}

\label{sec:app}

\subsection{Computational details}
	In all calculations the initial state is taken to be an $N_d$-dimensional Gaussian wave packet constructed as the product of one-dimensional Gaussians  and populated on a single potential energy surface.  The sinc pseudospectral method is used throughout as the reference.\cite{orszag:1969,gottlieb1977numerical,furnaro:1992,fornberg1998practical,Canuto:2006,hesthaven2007spectral, Lill1982483,Colbert:1992,Light:2000, Mazziotti1999473,Mazziotti:2002SD}  In the reference calculations, the propagator is calculated by diagonalizing and exponentiating the Hamiltonian operator followed by repeated application of the propagator.
	
The width $\alpha_{j}$ of each Gaussian basis function is set to the width of the initial state.  \textcolor{black}{The accuracy for either the spectral or pseudospectral versions of the Gaussian dynamics is not too sensitive to the choice of $\alpha_{j}$.  We have found in the time-independent limit that the pseudospectral Gaussian approximation favors a slightly broader Gaussian than the spectral Gaussian approximation.}~\cite{Heaps:2016} 	The initial position and momenta were sampled from the appropriate Wigner distribution of a Gaussian wavefunction.\cite{PhysRev.40.749, Heller:1976a}  The threshold to retain singular values was generally set to $1\times 10^{-4}$.  The equations were propagated using a fixed time step fourth-order Runge-Kutta algorithm.  The initial expansion coefficients for time-dependent problems are determined by projecting the basis onto the initial wavefunction,  $c(t=0) = \mathbf{\Phi}^{-1} \langle \delta(x-x_i)|\Psi\rangle$. The vector of elements $\langle \delta(x-x_i)|\Psi\rangle$, is the initial wavefunction evaluated at the basis function centers determined from sampling the Wigner distribution and $\mathbf{\Phi}^{-1}$ is the inverse of the discrete overlap matrix.  Although we employ the pseudospectral representation for propagation, expectation values on surfaces are calculated in the usual fashion, i.e.
		\begin{equation}
			P_I(t) = \langle \Phi^I(x,t) | \Phi^I(x,t) \rangle = (\mathbf{D}^I)^\dagger \mathbf{S} \mathbf{D}^I,
		\end{equation}
	where the overlap matrix, $\mathbf{S}$, is calculated analytically.
	
\subsection{Results}
\subsubsection{Morse potential}
	\begin{figure}[htb]
		  \centering
		  \includegraphics[width=0.48\textwidth]{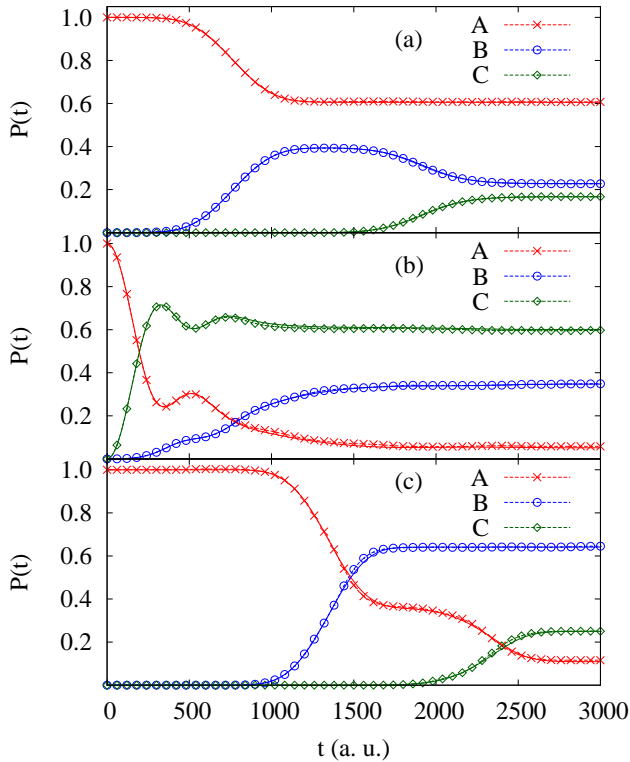}
		  \caption{The time-dependent populations for three coupled surfaces in a prototypical photodissociation process.  The initially occupied state (A) is denoted by x's, the second surface (B) by hollow circles, and the third (C) by hollow diamonds.  The exact results are given by solid lines and nearly indistinguishable for all cases.  All simulations used 150 trajectories.}
		  \label{fig:morse_3surf_pops}
	\end{figure}

	The first set of models investigated consists of Morse oscillator potential energy surfaces with Gaussian coupling between the surfaces.  The coupled Morse potentials are a  prototypical model for anharmonic vibrational dynamics and electron transfer.\cite{Coronado2001521, Bonella:2003, :/content/aip/journal/jcp/124/4/10.1063/1.2162172, doi:10.1080/00268976.2012.684896}  We will consider both  dissociative and bound state conditions.  The functional forms of the diabatic surfaces and coupling potentials are, respectively,
		\begin{equation}
			V_{ii}(x) = D_i\left(1-e^{-a_i(x-b_i)}\right)^2 + E_i
		\end{equation}
		\begin{equation}
			V_{ij}(x)=A_{ij}e^{-c_{ij}(x-d_{ij})^2}.
		\end{equation}

	First, we consider the photodissociation of a wave packet in a system of three coupled potential energy surfaces.\cite{Coronado2001521,doi:10.1080/00268976.2012.684896, doi:10.1021/acs.jpclett.5b01957}  The original system parameters for the three cases may be found in Ref.~\onlinecite{Coronado2001521}.  The calculation is meant to model photodissociation following excitation from a harmonic ground state.  The initial wave packet is taken to be high on the repulsive barrier, leading to dissociation after passage through the regions of nonadiabatic coupling.  Given the qualitative similarity of the three Morse potentials, we can expect that Ehrenfest trajectories will appropriately cover important regions of phase space.  Each simulation used 150 trajectories, a time step of 3 a.u., and an SVD threshold of $1\times 10^{-4}$.
	
	The time-dependent populations for the three states are presented in Fig.~\ref{fig:morse_3surf_pops}.  In all three cases, 150 trajectories are sufficient to produce results indistinguishable from the exact calculation.  In the first and third cases, the two regions of coupling are well-separated spatially while the second model couples all three surfaces in proximity.  In all cases the pseudospectral Gaussian method correctly predicts the population exchange between all three surfaces.  Since the pseudospectral Gaussian method solves the Schr\"odinger equation in matrix form, one might expect the proper treatment of the population transfer.  We demonstrate that one may obtain accurate and efficient solutions by projecting the disordered Gaussian basis set onto discrete points in space.
		
	A more challenging case is the bound-state dynamics of two coupled Morse potentials $A$ and $B$.  In this model, the photo excitation results in a Gaussian wave packet starting on the shallow, attractive region of the initially occupied state, leading to oscillatory dynamics and many crossings through the region of nonadiabatic coupling.  The parameters used in this paper are $D_A=2.278\times10^{-2}$, $a_A = 0.675$, $b_A = 1.89$, $E_A = 0.0$.  $D_B = 1.025\times 10^{-2}$, $a_B = 0.453$, $b_B = 3.212$, $E_B = 3.8\times10^{-3}$, $d_{AB} = 2.744$, $c_{AB} = 0.56$, $A_{AB} = 6.337\times10^{-3}$.  They are slightly modified from the work of Coker and co-workers.\cite{Bonella:2003, doi:10.1080/00268976.2012.684896}  The initial wave packet parameters are $x_c=4.0$ a.u., $k_0 = 0.0$ a.u., mass$= 2000.0$ a.u. and $\alpha = 0.5$ a.u.$^{-2}$.  The simulation is run for 10,000 a.u., approximately 240 femtoseconds, a time step of 5 a.u.\ and an SVD threshold of $1\times 10^{-2}$.\cite{doi:10.1080/00268976.2012.684896}
	
	\begin{figure}[!htb]
		  \centering
		  \includegraphics[width=0.48\textwidth]{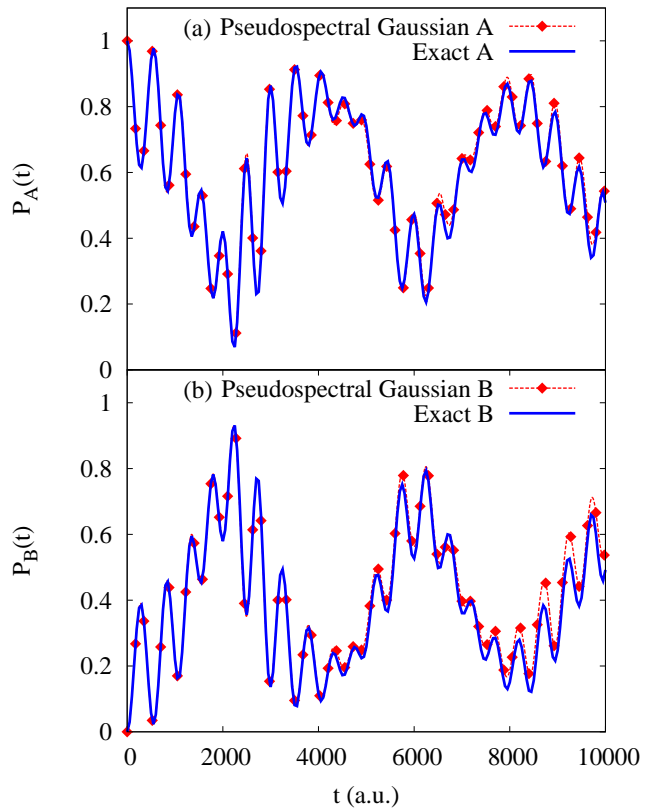}
		  \caption{Time-dependent populations of the diabatic states for the bound state Morse model surface A (a) and B (b) using 1,000 trajectories.  Both the high and low frequency oscillations corresponding to continuous nonadiabatic exchange and nuclear motion, respectively, are reproduced.}
		  \label{fig:morse_bound_pops_both}
	\end{figure}	
	\begin{figure}[!htb]
		  \centering
		  \includegraphics[width=0.48\textwidth]{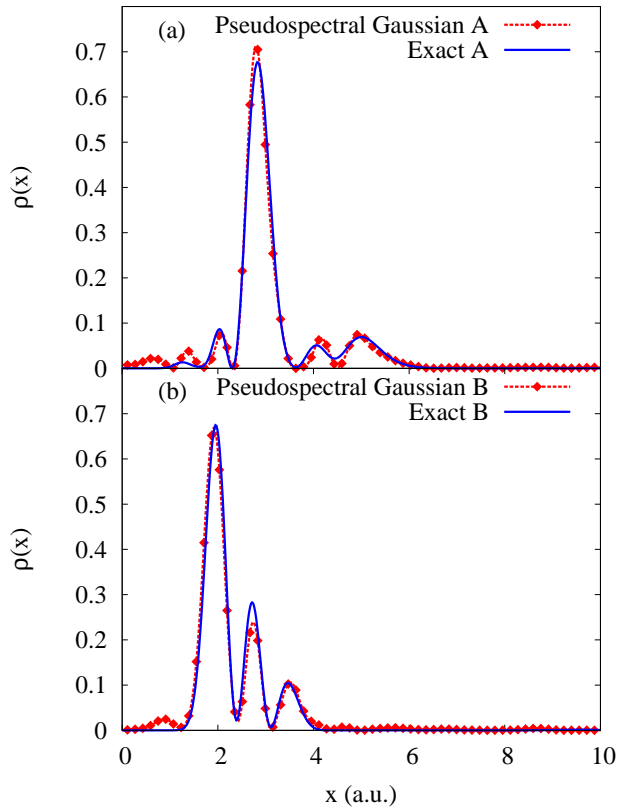}
		  \caption{The density on surface A (a) and surface B (b) of the bound state Morse model at $t = 10,000$ a.u.\ using 1,000 trajectories.  The pseudospectral Gaussian captures the nodal features in the density characteristic of coherent quantum dynamics in a Morse potential.}
		  \label{fig:morse_bound_dens_both}
	\end{figure}
	
	\begin{figure}[!htb]
		  \centering
		  \includegraphics[width=0.48\textwidth]{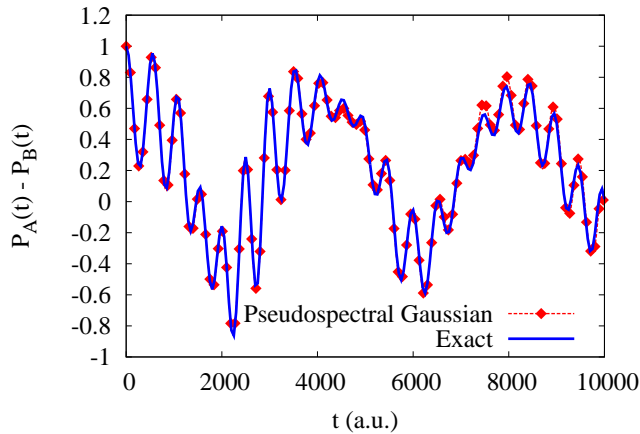}
		  \caption{Time-dependent population difference of the diabatic surfaces for the bound state Morse potential using 250 trajectories.  While the population differences quantitatively agree, calculating the population difference obscures the deviation in total norm.  Nevertheless, the agreement is excellent.}
		  \label{fig:morse_bound_diff_small_basis}
	\end{figure}
	
	The time-dependent population dynamics using 1,000 trajectories are presented in Fig.~\ref{fig:morse_bound_pops_both}.  The bound state is characterized by two important time scales.  First, there is the high frequency population exchange caused by continuous nonadiabatic transfer.  There is also a slower oscillation in the populations corresponding to the nuclear wave packet motion.  As a result, accurate population dynamics requires proper treatment of both inter- and intra-state coupling.  The pseudospectral Gaussian method properly captures both of these effects, leading to quantitative agreement over the entire propagation.  As a test of the quality of the intra-surface coupling, we plot the densities for the two surfaces at the final time in Fig.~\ref{fig:morse_bound_dens_both}.  The nodal features characteristic of the coherent dynamics are reproduced very well by the pseudospectral sampling.	
	
	While the results are well converged for 1,000 trajectories, we also present results using 250 trajectories in Fig.~\ref{fig:morse_bound_diff_small_basis} where the population difference, $P_A - P_B$, rather than the populations is presented.  The much smaller basis of trajectories quantitatively describes the population exchange between the two states.  However, the total norm of the system at $t=10,000$ is 1.23.  The deviation in norm reflects the breakdown in the method as the basis set no longer sufficiently covers the important regions of phase space.  Despite the accumulated error in the total norm, the pseudospectral Gaussian method still offers a quantitative description of the population exchange.
	
\subsubsection{Two-Dimensional Conical Intersection}
		\begin{figure}[!htb]
		  \centering
		  \includegraphics[width=0.48\textwidth]{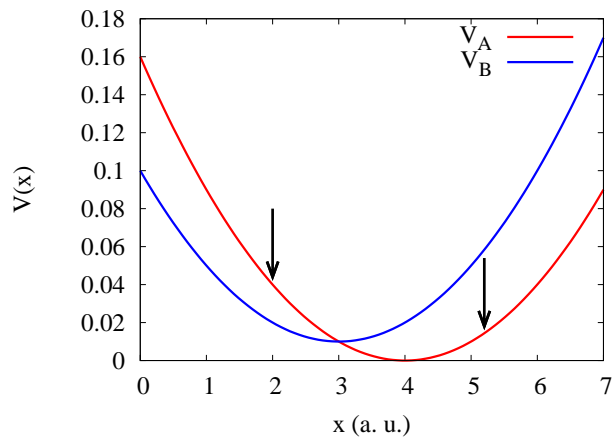}
		  \caption{A cross section of the two-dimensional diabatic potentials along the x-axis at $y=0$.  The arrows mark the two initial wave packet positions.   Both wave packets begin on V$_{\mathrm{A}}$.}
		  \label{fig:ci2d_potential}
	\end{figure}

Finally, we investigate the nonadiabatic dynamics of a model for the vibrational dynamics of a collinear triatomic molecule $ABA$.\cite{Ferretti:1996,:/content/aip/journal/jcp/130/13/10.1063/1.3103930}  The model describes the coupling between two electronic states with two degrees of freedom, the symmetric ($x$) and anti-symmetric ($y$) vibrational modes.  The potential energy surfaces are
	\begin{align}
		V_{A}(x,y) &= \frac{1}{2}k_x(x-x_1)^2 + \frac{1}{2} k_y y^2 \nonumber \\
		V_{B}(x,y) &= \frac{1}{2}k_x(x-x_2)^2 + \frac{1}{2} k_y y^2 + \Delta \\
		V_{C}(x,y) &= \gamma y \exp\left(-\alpha(x-x_3)^2 -\beta y^2 \right)
	\end{align}
	where $x_1 = 4$, $x_2 = x_3 = 3$, $k_x = 0.02$, $k_y= 0.1$, $\Delta = 0.01$, $\gamma = 0.01$, $\alpha = 3$ and $\beta = 1.5$.  The parameter $\gamma$ controls the interstate coupling for the model.  The initial wave packet is selected to model a Franck-Condon excitation from a harmonic ground state.  The masses are $m_x = 20 000, m_y = 6667$ a.u., the initial wave packet widths are $\alpha_x = 22.2$ and $\alpha_y = 12.9$ a.u.$^{-2}$, and the wave packet is centered at $y_0=0$ for both examples.  In both cases $p_x = p_y = 0$.
		\begin{figure}[!htb]
		  \centering
		  \includegraphics[width=0.48\textwidth]{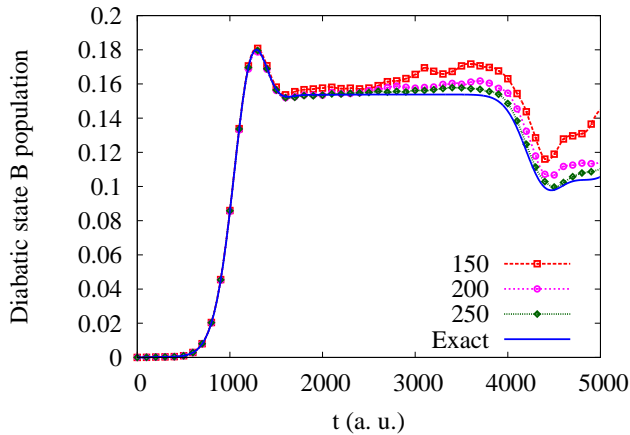}
		  \caption{Time-dependent population for the upper diabatic surface for increasing basis set size and an initial wave packet centered at $(x,y) = (2, 0)$.  The initial condition leads to a higher energy wave packet that completes the first passage through the region of nonadiabatic coupling at approximately 1,200 a.u.  The three basis set sizes exhibit clear convergence to the exact solution with excellent agreement using 250 trajectories.  }
		  \label{fig:ci2d_2_pops}
	\end{figure}
	\begin{figure}[!htb]
		  \centering
		  \includegraphics[width=0.48\textwidth]{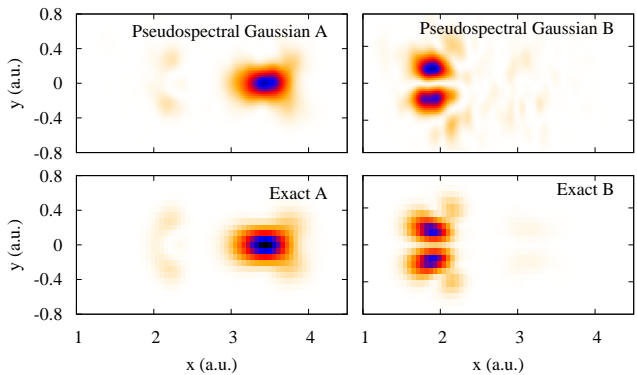}
		  \caption{The wave packet densities at $t = 5,000$ a.u.\ from the wave packet starting at  $(x,y) = (2, 0)$.  The upper panels are the pseudospectral Gaussian method and the lower panels the exact grid calculation.   There are patches of spurious density on the upper surface but otherwise all of the features are reproduced very well.   Note the spatial separation of the density on the two surfaces.}
		  \label{fig:ci2d_2_dens}
	\end{figure}

	First, we consider a wave packet starting at $x _0= 2.0$ on surface A.  As shown in Fig.~\ref{fig:ci2d_potential}, the initial condition corresponds to an energy well above the crossing region.  The wave packet was propagated for 5,000 a.u., capturing the initial passage through the coupling region and a second period where the density in the excited state returns to the coupling region, leading to a small amount of population transfer back to the ground state.  All basis set sizes predict the initial population transfer in excellent agreement with the exact method.  However, as the simulation progresses, the smaller basis sets deteriorate in quality even when outside the region of nonadiabatic coupling.  Inspection of the trajectories suggest the spurious population, accompanied by deviation of total norm, occurs when the Ehrenfest trajectories no longer cover the regions of density on the upper surface.
	
	\begin{figure}[!htb]
		  \centering
		  \includegraphics[width=0.48\textwidth]{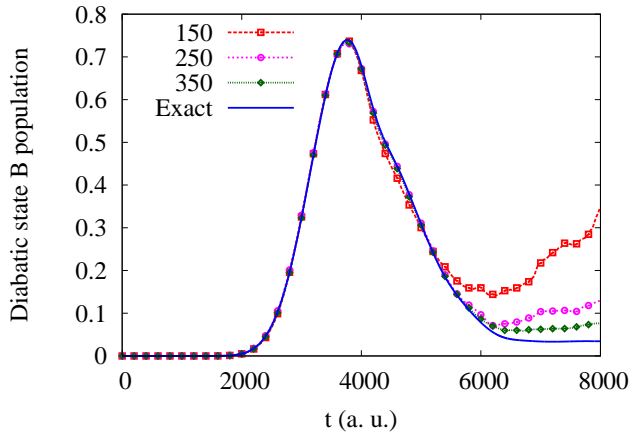}
		  \caption{Time-dependent population for the upper diabatic surface for increasing basis set size and an initial wave packet centered at $(x,y) = (5.2, 0)$.  The initial energy of the wave packet is approximately equal to the energy of the surface crossing resulting in many trajectories not reaching the crossing region.  While there is significant population exchange upon reaching the crossing point, most of the density returns to the ground state following reflection off of the harmonic barrier.  The smaller basis set predicts spurious population on the upper surface, particularly as the densities spatially separate and the Ehrenfest trajectories fail.  The error is greatly reduced as basis set size increases.}
		  \label{fig:ci2d_5_pops}
	\end{figure}
	The breakdown of Ehrenfest trajectories and the qualitative features of population transfer may be better understood by inspecting the densities on the respective surfaces at $t = 5,000$ a.u.\ in Fig.~\ref{fig:ci2d_2_dens}.  Since the wave packet begins at $(x,y) = (2,0)$, it first moves in the positive $x$-direction and passes through the region of derivative coupling, completing the initial population transfer in the first 1,500 time units.  Now, a small portion of the wave packet is propagating on the upper surface while most remains in the ground state.  However, the classical turning point on surface B is at $x = 4.7$ but on surface A it is at $x=6.0$.  Therefore the momentum of the density on B changes sign earlier in time than on A and passes through the region of derivative coupling again at 4,000 a.u.  Since the wave packet is still predominantly on the lower surface, one would expect the trajectories to follow the forces of that surface.  This, of course, prevents trajectories from following the upper surface through the crossing at $t = 4,000$ a.u.  The densities illustrate the spatial separation caused by the different forces.
	
	The second crossing transfers some of the population from B back to A.  This density is observable in both the approximate and exact calculations on surface A at $x = 2$, spatially separated from the principal wave packet density.  The pseudospectral Gaussian method also reproduces the node in the upper surface that is a consequence of the Berry phase.\cite{Longuet-Higgins147,Berry45}  This phase relationship is only observable from a proper quantum-mechanical treatment of the dynamics.    While the qualitative failure of Ehrenfest trajectories is overcome by using larger basis sets, basis function spawning is a much more efficient solution that may be pursued in the future.
	
	In the second example, we begin with a wave packet centered at $x_0 = 5.2$, corresponding to lower energy dynamics.  In this case, the wave packet reaches the crossing region with very little excess energy and many trajectories will not reach the intersection.  As discussed in  Yang \emph{et al.\ }, this type of transition is very difficult for surface hopping methods.\cite{:/content/aip/journal/jcp/130/13/10.1063/1.3103930}  The time-dependent probability on the excited surface is plotted in Fig.~\ref{fig:ci2d_5_pops}.  Similar to the first case, excellent agreement is observed for all basis set sizes for the first half of the propagation.  At this time, the densities on the two states have considerably different momenta and the trajectories fail to follow the quantum mechanics.  Interestingly, the pseudospectral Gaussian method converges to a final excited state population slightly above the exact result.  While improved accuracy would be preferred, we note that the Gaussian based methods in Ref.\ \cite{:/content/aip/journal/jcp/130/13/10.1063/1.3103930}
 converge to a similar population.

\section{Discussion}	
	In this work, we introduce a Gaussian trajectory based approach to non-adiabatic dynamics that  only requires $\mathcal{O}(N)$ potential energy calculations yet describes quantum-mechanical coherence in the nuclear dynamics.  Selecting the Dirac delta function to test the Schr\"odinger equation produces an efficient set of equations that circumvents the costly and inaccurate numerical integration of the potential energy generally associated with Gaussian basis sets.   While traditional pseudospectral methods require highly structured grids, we demonstrate that  accurate dynamics may still be realized despite abandoning a structured grid and basis function orthogonality.
	
The trajectory-guided basis of the pseudospectral Gaussian method connects the method to many other trajectory-based methods.  Unlike surface hopping and semi-classical methods,\cite{tully:1061, Webster1991494,hammes-schiffer:4657,Prezhdo:1997,PhysRevLett.78.578,sun:6346,A801824C,Bonella:2003,Wu:2005kx,doi:10.1021/jp809907p,:/content/aip/journal/jcp/135/20/10.1063/1.3664763,doi:10.1080/00268976.2012.684896,:/content/aip/journal/jcp/141/8/10.1063/1.4893345}  quantum mechanics is incorporated explicitly by solving the Schr\"odinger equation in matrix form at each time step.  The unstructured grid-like framework of pseudospectral Gaussian dynamics is adopted from Bohmian mechanics, where quantum-mechanical amplitudes are propagated at discrete points in space.  However, the matrix formulation is fundamentally distinct from Bohmian mechanics and does not suffer from the numerical instabilities associated with Bohmian mechanics.\cite{PhysRevLett.82.5190,PhysRevE.61.5967,kendrick:5805,wyatt2006quantum}
	
	Effective dynamics using a trajectory-guided basis set relies on two components; efficient, local description of the quantum mechanics and coverage of important regions in phase space.  In this work, we focus on the former, using the pseudospectral Gaussian method to solve the quantum mechanics.  The results in this paper suggest that, when the basis set properly reflects the quantum mechanics, the pseudospectral Gaussian method is very accurate.  However, the Ehrenfest trajectories are suboptimal, particularly in the two-dimensional model presented here where the displacement of the surfaces lead to substantially different gradients for populations on the respective surfaces.  A future direction lies in coupling the pseudospectral Gaussian method with a surface hopping or a spawning procedure, both of which improve upon the limitations of Ehrenfest trajectories.

	Employing pseudospectral sampling offers a promising new approach to Gaussian-based dynamics.  The method adopts many of the attractive features of moving Gaussian basis sets while circumventing one of their greatest difficulties, the potential energy integral evaluation.  We demonstrate that the pseudospectral Gaussian dynamics accurately describes both dissociative and bound-state processes using a coupled Morse potential model and a model for collinear triatomic vibration.  Using Ehrenfest guided trajectories in the method, we are able to describe simultaneously population dynamics and intra-surface dynamics for wave packets for long times.  The successful implementation of the pseudospectral Gaussian method to one- and two-dimensional nonadiabatic models suggests the method may be well-suited for \emph{ab initio} on-the-fly non-adiabatic quantum molecular dynamics.

\begin{acknowledgments}
D.A.M. gratefully acknowledges the U.S. National Science Foundation CHE-1565638 the U.S. Army Research Office (ARO)  Grant No. W911NF-16-1-0152 and W911NF-16-C-0030, and the U.S. Air Force Office of Scientific Research (AFOSR) FA9550-14-1-0367 for their support.  C.W.H.~gratefully acknowledges financial support from the Department of Education through the Graduate Assistance in Areas of National Need Fellowship (Grant No. P200A120093).
\end{acknowledgments}
	
\bibliography{nonadiabatic_paper_refs_R3}

\begin{thebibliography}{112}%
\makeatletter
\providecommand \@ifxundefined [1]{%
 \@ifx{#1\undefined}
}%
\providecommand \@ifnum [1]{%
 \ifnum #1\expandafter \@firstoftwo
 \else \expandafter \@secondoftwo
 \fi
}%
\providecommand \@ifx [1]{%
 \ifx #1\expandafter \@firstoftwo
 \else \expandafter \@secondoftwo
 \fi
}%
\providecommand \natexlab [1]{#1}%
\providecommand \enquote  [1]{``#1''}%
\providecommand \bibnamefont  [1]{#1}%
\providecommand \bibfnamefont [1]{#1}%
\providecommand \citenamefont [1]{#1}%
\providecommand \href@noop [0]{\@secondoftwo}%
\providecommand \href [0]{\begingroup \@sanitize@url \@href}%
\providecommand \@href[1]{\@@startlink{#1}\@@href}%
\providecommand \@@href[1]{\endgroup#1\@@endlink}%
\providecommand \@sanitize@url [0]{\catcode `\\12\catcode `\$12\catcode
  `\&12\catcode `\#12\catcode `\^12\catcode `\_12\catcode `\%12\relax}%
\providecommand \@@startlink[1]{}%
\providecommand \@@endlink[0]{}%
\providecommand \url  [0]{\begingroup\@sanitize@url \@url }%
\providecommand \@url [1]{\endgroup\@href {#1}{\urlprefix }}%
\providecommand \urlprefix  [0]{URL }%
\providecommand \Eprint [0]{\href }%
\providecommand \doibase [0]{http://dx.doi.org/}%
\providecommand \selectlanguage [0]{\@gobble}%
\providecommand \bibinfo  [0]{\@secondoftwo}%
\providecommand \bibfield  [0]{\@secondoftwo}%
\providecommand \translation [1]{[#1]}%
\providecommand \BibitemOpen [0]{}%
\providecommand \bibitemStop [0]{}%
\providecommand \bibitemNoStop [0]{.\EOS\space}%
\providecommand \EOS [0]{\spacefactor3000\relax}%
\providecommand \BibitemShut  [1]{\csname bibitem#1\endcsname}%
\let\auto@bib@innerbib\@empty
\bibitem [{\citenamefont {Yarkony}(1996)}]{doi:10.1021/jp962134y}%
  \BibitemOpen
  \bibfield  {author} {\bibinfo {author} {\bibfnamefont {D.~R.}\ \bibnamefont
  {Yarkony}},\ }\href@noop {} {\bibfield  {journal} {\bibinfo  {journal} {J.
  Phys. Chem.}\ }\textbf {\bibinfo {volume} {100}},\ \bibinfo {pages} {18612}
  (\bibinfo {year} {1996})}\BibitemShut {NoStop}%
\bibitem [{\citenamefont
  {Butler}(1998)}]{doi:10.1146/annurev.physchem.49.1.125}%
  \BibitemOpen
  \bibfield  {author} {\bibinfo {author} {\bibfnamefont {L.~J.}\ \bibnamefont
  {Butler}},\ }\href@noop {} {\bibfield  {journal} {\bibinfo  {journal} {Ann.
  Rev. Phys. Chem.}\ }\textbf {\bibinfo {volume} {49}},\ \bibinfo {pages} {125}
  (\bibinfo {year} {1998})}\BibitemShut {NoStop}%
\bibitem [{\citenamefont {Tully}(2012)}]{tully:22A301}%
  \BibitemOpen
  \bibfield  {author} {\bibinfo {author} {\bibfnamefont {J.~C.}\ \bibnamefont
  {Tully}},\ }\href@noop {} {\bibfield  {journal} {\bibinfo  {journal} {J.
  Chem. Phys.}\ }\textbf {\bibinfo {volume} {137}},\ \bibinfo {eid} {22A301}
  (\bibinfo {year} {2012})}\BibitemShut {NoStop}%
\bibitem [{\citenamefont {Persico}\ and\ \citenamefont
  {Granucci}(2014)}]{persico:2014}%
  \BibitemOpen
  \bibfield  {author} {\bibinfo {author} {\bibfnamefont {M.}~\bibnamefont
  {Persico}}\ and\ \bibinfo {author} {\bibfnamefont {G.}~\bibnamefont
  {Granucci}},\ }\href@noop {} {\bibfield  {journal} {\bibinfo  {journal}
  {Theo. Chem. Acc.}\ }\textbf {\bibinfo {volume} {133}} (\bibinfo {year}
  {2014})}\BibitemShut {NoStop}%
\bibitem [{\citenamefont {Tavernelli}(2015)}]{doi:10.1021/ar500357y}%
  \BibitemOpen
  \bibfield  {author} {\bibinfo {author} {\bibfnamefont {I.}~\bibnamefont
  {Tavernelli}},\ }\href@noop {} {\bibfield  {journal} {\bibinfo  {journal}
  {Acc. Chem. Res.}\ }\textbf {\bibinfo {volume} {48}},\ \bibinfo {pages} {792}
  (\bibinfo {year} {2015})}\BibitemShut {NoStop}%
\bibitem [{\citenamefont {Ben-Nun}\ and\ \citenamefont
  {Mart{\'\i}nez}(1998{\natexlab{a}})}]{BenNun199857}%
  \BibitemOpen
  \bibfield  {author} {\bibinfo {author} {\bibfnamefont {M.}~\bibnamefont
  {Ben-Nun}}\ and\ \bibinfo {author} {\bibfnamefont {T.~J.}\ \bibnamefont
  {Mart{\'\i}nez}},\ }\href@noop {} {\bibfield  {journal} {\bibinfo  {journal}
  {Chem. Phys. Lett.}\ }\textbf {\bibinfo {volume} {298}},\ \bibinfo {pages}
  {57 } (\bibinfo {year} {1998}{\natexlab{a}})}\BibitemShut {NoStop}%
\bibitem [{\citenamefont {Ben-Nun}\ and\ \citenamefont
  {Mart{\'\i}nez}(1999)}]{BenNun:1999a}%
  \BibitemOpen
  \bibfield  {author} {\bibinfo {author} {\bibfnamefont {M.}~\bibnamefont
  {Ben-Nun}}\ and\ \bibinfo {author} {\bibfnamefont {T.~J.}\ \bibnamefont
  {Mart{\'\i}nez}},\ }\href@noop {} {\bibfield  {journal} {\bibinfo  {journal}
  {J. Chem. Phys.}\ }\textbf {\bibinfo {volume} {110}},\ \bibinfo {pages}
  {4134} (\bibinfo {year} {1999})}\BibitemShut {NoStop}%
\bibitem [{\citenamefont {Ben-Nun}, \citenamefont {Quenneville},\ and\
  \citenamefont {Mart{\'\i}nez}(2000)}]{doi:10.1021/jp994174i}%
  \BibitemOpen
  \bibfield  {author} {\bibinfo {author} {\bibfnamefont {M.}~\bibnamefont
  {Ben-Nun}}, \bibinfo {author} {\bibfnamefont {J.}~\bibnamefont
  {Quenneville}}, \ and\ \bibinfo {author} {\bibfnamefont {T.~J.}\ \bibnamefont
  {Mart{\'\i}nez}},\ }\href@noop {} {\bibfield  {journal} {\bibinfo  {journal}
  {J. Phys. Chem. A}\ }\textbf {\bibinfo {volume} {104}},\ \bibinfo {pages}
  {5161} (\bibinfo {year} {2000})}\BibitemShut {NoStop}%
\bibitem [{\citenamefont {Ben-Nun}\ and\ \citenamefont
  {Mart{\'\i}nez}(2002)}]{ben-nun:2002}%
  \BibitemOpen
  \bibfield  {author} {\bibinfo {author} {\bibfnamefont {M.}~\bibnamefont
  {Ben-Nun}}\ and\ \bibinfo {author} {\bibfnamefont {T.~J.}\ \bibnamefont
  {Mart{\'\i}nez}},\ }\href@noop {} {\bibfield  {journal} {\bibinfo  {journal}
  {Adv. Chem. Phys.}\ }\textbf {\bibinfo {volume} {121}},\ \bibinfo {pages}
  {439} (\bibinfo {year} {2002})}\BibitemShut {NoStop}%
\bibitem [{\citenamefont {Iyengar}\ and\ \citenamefont
  {Jakowski}(2005)}]{:/content/aip/journal/jcp/122/11/10.1063/1.1871876}%
  \BibitemOpen
  \bibfield  {author} {\bibinfo {author} {\bibfnamefont {S.~S.}\ \bibnamefont
  {Iyengar}}\ and\ \bibinfo {author} {\bibfnamefont {J.}~\bibnamefont
  {Jakowski}},\ }\href@noop {} {\bibfield  {journal} {\bibinfo  {journal} {J.
  Chem. Phys.}\ }\textbf {\bibinfo {volume} {122}},\ \bibinfo {pages} {114105}
  (\bibinfo {year} {2005})}\BibitemShut {NoStop}%
\bibitem [{\citenamefont {Lasorne}\ \emph {et~al.}(2006)\citenamefont
  {Lasorne}, \citenamefont {Bearpark}, \citenamefont {Robb},\ and\
  \citenamefont {Worth}}]{Lasorne2006604}%
  \BibitemOpen
  \bibfield  {author} {\bibinfo {author} {\bibfnamefont {B.}~\bibnamefont
  {Lasorne}}, \bibinfo {author} {\bibfnamefont {M.~J.}\ \bibnamefont
  {Bearpark}}, \bibinfo {author} {\bibfnamefont {M.~A.}\ \bibnamefont {Robb}},
  \ and\ \bibinfo {author} {\bibfnamefont {G.~A.}\ \bibnamefont {Worth}},\
  }\href@noop {} {\bibfield  {journal} {\bibinfo  {journal} {Chem. Phys.
  Lett.}\ }\textbf {\bibinfo {volume} {432}},\ \bibinfo {pages} {604 }
  (\bibinfo {year} {2006})}\BibitemShut {NoStop}%
\bibitem [{\citenamefont {Lasorne}, \citenamefont {Robb},\ and\ \citenamefont
  {Worth}(2007)}]{B700297A}%
  \BibitemOpen
  \bibfield  {author} {\bibinfo {author} {\bibfnamefont {B.}~\bibnamefont
  {Lasorne}}, \bibinfo {author} {\bibfnamefont {M.~A.}\ \bibnamefont {Robb}}, \
  and\ \bibinfo {author} {\bibfnamefont {G.~A.}\ \bibnamefont {Worth}},\
  }\href@noop {} {\bibfield  {journal} {\bibinfo  {journal} {Phys. Chem. Chem.
  Phys.}\ }\textbf {\bibinfo {volume} {9}},\ \bibinfo {pages} {3210} (\bibinfo
  {year} {2007})}\BibitemShut {NoStop}%
\bibitem [{\citenamefont {Li}\ and\ \citenamefont
  {Iyengar}(2010)}]{:/content/aip/journal/jcp/133/18/10.1063/1.3504167}%
  \BibitemOpen
  \bibfield  {author} {\bibinfo {author} {\bibfnamefont {X.}~\bibnamefont
  {Li}}\ and\ \bibinfo {author} {\bibfnamefont {S.~S.}\ \bibnamefont
  {Iyengar}},\ }\href@noop {} {\bibfield  {journal} {\bibinfo  {journal} {J.
  Chem. Phys.}\ }\textbf {\bibinfo {volume} {133}},\ \bibinfo {pages} {184105}
  (\bibinfo {year} {2010})}\BibitemShut {NoStop}%
\bibitem [{\citenamefont {Saita}\ and\ \citenamefont
  {Shalashilin}(2012)}]{:/content/aip/journal/jcp/137/22/10.1063/1.4734313}%
  \BibitemOpen
  \bibfield  {author} {\bibinfo {author} {\bibfnamefont {K.}~\bibnamefont
  {Saita}}\ and\ \bibinfo {author} {\bibfnamefont {D.~V.}\ \bibnamefont
  {Shalashilin}},\ }\href@noop {} {\bibfield  {journal} {\bibinfo  {journal}
  {J. Chem. Phys.}\ }\textbf {\bibinfo {volume} {137}},\ \bibinfo {pages}
  {22A506} (\bibinfo {year} {2012})}\BibitemShut {NoStop}%
\bibitem [{\citenamefont {Leveque}\ \emph {et~al.}(2013)\citenamefont
  {Leveque}, \citenamefont {Komainda}, \citenamefont {Taieb},\ and\
  \citenamefont {Koppel}}]{leveque:044320}%
  \BibitemOpen
  \bibfield  {author} {\bibinfo {author} {\bibfnamefont {C.}~\bibnamefont
  {Leveque}}, \bibinfo {author} {\bibfnamefont {A.}~\bibnamefont {Komainda}},
  \bibinfo {author} {\bibfnamefont {R.}~\bibnamefont {Taieb}}, \ and\ \bibinfo
  {author} {\bibfnamefont {H.}~\bibnamefont {Koppel}},\ }\href@noop {}
  {\bibfield  {journal} {\bibinfo  {journal} {J. Chem. Phys.}\ }\textbf
  {\bibinfo {volume} {138}},\ \bibinfo {eid} {044320} (\bibinfo {year}
  {2013})}\BibitemShut {NoStop}%
\bibitem [{\citenamefont {Makhov}\ \emph {et~al.}(2014)\citenamefont {Makhov},
  \citenamefont {Glover}, \citenamefont {Mart{\'\i}nez},\ and\ \citenamefont
  {Shalashilin}}]{Makhov:2014}%
  \BibitemOpen
  \bibfield  {author} {\bibinfo {author} {\bibfnamefont {D.~V.}\ \bibnamefont
  {Makhov}}, \bibinfo {author} {\bibfnamefont {W.~J.}\ \bibnamefont {Glover}},
  \bibinfo {author} {\bibfnamefont {T.~J.}\ \bibnamefont {Mart{\'\i}nez}}, \
  and\ \bibinfo {author} {\bibfnamefont {D.~V.}\ \bibnamefont {Shalashilin}},\
  }\href@noop {} {\bibfield  {journal} {\bibinfo  {journal} {J. Chem. Phys.}\
  }\textbf {\bibinfo {volume} {141}},\ \bibinfo {pages} {054110} (\bibinfo
  {year} {2014})}\BibitemShut {NoStop}%
\bibitem [{\citenamefont {Lasorne}, \citenamefont {Worth},\ and\ \citenamefont
  {Robb}(2014)}]{lasorne:2014}%
  \BibitemOpen
  \bibfield  {author} {\bibinfo {author} {\bibfnamefont {B.}~\bibnamefont
  {Lasorne}}, \bibinfo {author} {\bibfnamefont {G.}~\bibnamefont {Worth}}, \
  and\ \bibinfo {author} {\bibfnamefont {M.}~\bibnamefont {Robb}},\ }in\
  \href@noop {} {\emph {\bibinfo {booktitle} {Molecular Quantum Dynamics}}},\
  \bibinfo {editor} {edited by\ \bibinfo {editor} {\bibfnamefont
  {F.}~\bibnamefont {Gatti}}}\ (\bibinfo  {publisher} {Springer Berlin
  Heidelberg},\ \bibinfo {year} {2014})\ pp.\ \bibinfo {pages}
  {181--211}\BibitemShut {NoStop}%
\bibitem [{\citenamefont {Roos}(1999)}]{doi:10.1021/ar960091y}%
  \BibitemOpen
  \bibfield  {author} {\bibinfo {author} {\bibfnamefont {B.~O.}\ \bibnamefont
  {Roos}},\ }\href@noop {} {\bibfield  {journal} {\bibinfo  {journal} {Acc.
  Chem. Res.}\ }\textbf {\bibinfo {volume} {32}},\ \bibinfo {pages} {137}
  (\bibinfo {year} {1999})}\BibitemShut {NoStop}%
\bibitem [{\citenamefont {Coe}\ \emph {et~al.}(2007)\citenamefont {Coe},
  \citenamefont {Levine}, ,\ and\ \citenamefont
  {Mart{\'\i}nez}}]{doi:10.1021/jp072027b}%
  \BibitemOpen
  \bibfield  {author} {\bibinfo {author} {\bibfnamefont {J.~D.}\ \bibnamefont
  {Coe}}, \bibinfo {author} {\bibfnamefont {B.~G.}\ \bibnamefont {Levine}}, , \
  and\ \bibinfo {author} {\bibfnamefont {T.~J.}\ \bibnamefont
  {Mart{\'\i}nez}},\ }\href@noop {} {\bibfield  {journal} {\bibinfo  {journal}
  {J. Phys. Chem. A}\ }\textbf {\bibinfo {volume} {111}},\ \bibinfo {pages}
  {11302} (\bibinfo {year} {2007})}\BibitemShut {NoStop}%
\bibitem [{\citenamefont {Tsuchimochi}\ and\ \citenamefont
  {Scuseria}(2009)}]{:/content/aip/journal/jcp/131/12/10.1063/1.3237029}%
  \BibitemOpen
  \bibfield  {author} {\bibinfo {author} {\bibfnamefont {T.}~\bibnamefont
  {Tsuchimochi}}\ and\ \bibinfo {author} {\bibfnamefont {G.~E.}\ \bibnamefont
  {Scuseria}},\ }\href@noop {} {\bibfield  {journal} {\bibinfo  {journal} {J.
  Chem. Phys.}\ }\textbf {\bibinfo {volume} {131}},\ \bibinfo {eid} {121102}
  (\bibinfo {year} {2009})}\BibitemShut {NoStop}%
\bibitem [{\citenamefont {Tao}, \citenamefont {Levine},\ and\ \citenamefont
  {Mart{\'\i}nez}(2009)}]{doi:10.1021/jp9063565}%
  \BibitemOpen
  \bibfield  {author} {\bibinfo {author} {\bibfnamefont {H.}~\bibnamefont
  {Tao}}, \bibinfo {author} {\bibfnamefont {B.~G.}\ \bibnamefont {Levine}}, \
  and\ \bibinfo {author} {\bibfnamefont {T.~J.}\ \bibnamefont
  {Mart{\'\i}nez}},\ }\href@noop {} {\bibfield  {journal} {\bibinfo  {journal}
  {J. Phys. Chem. A}\ }\textbf {\bibinfo {volume} {113}},\ \bibinfo {pages}
  {13656} (\bibinfo {year} {2009})}\BibitemShut {NoStop}%
\bibitem [{\citenamefont {Yanai}\ \emph {et~al.}(2010)\citenamefont {Yanai},
  \citenamefont {Kurashige}, \citenamefont {Neuscamman},\ and\ \citenamefont
  {Chan}}]{:/content/aip/journal/jcp/132/2/10.1063/1.3275806}%
  \BibitemOpen
  \bibfield  {author} {\bibinfo {author} {\bibfnamefont {T.}~\bibnamefont
  {Yanai}}, \bibinfo {author} {\bibfnamefont {Y.}~\bibnamefont {Kurashige}},
  \bibinfo {author} {\bibfnamefont {E.}~\bibnamefont {Neuscamman}}, \ and\
  \bibinfo {author} {\bibfnamefont {G.~K.-L.}\ \bibnamefont {Chan}},\
  }\href@noop {} {\bibfield  {journal} {\bibinfo  {journal} {J. Chem. Phys.}\
  }\textbf {\bibinfo {volume} {132}},\ \bibinfo {eid} {024105} (\bibinfo {year}
  {2010})}\BibitemShut {NoStop}%
\bibitem [{\citenamefont {Mazziotti}(2012)}]{Mazziotti2012}%
  \BibitemOpen
  \bibfield  {author} {\bibinfo {author} {\bibfnamefont {D.~A.}\ \bibnamefont
  {Mazziotti}},\ }\href {\doibase 10.1021/cr2000493} {\bibfield  {journal}
  {\bibinfo  {journal} {Chem. Rev.}\ }\textbf {\bibinfo {volume} {112}},\
  \bibinfo {pages} {244} (\bibinfo {year} {2012})}\BibitemShut {NoStop}%
\bibitem [{\citenamefont {Heaps}\ and\ \citenamefont
  {Mazziotti}(2016)}]{Heaps:2016}%
  \BibitemOpen
  \bibfield  {author} {\bibinfo {author} {\bibfnamefont {C.~W.}\ \bibnamefont
  {Heaps}}\ and\ \bibinfo {author} {\bibfnamefont {D.~A.}\ \bibnamefont
  {Mazziotti}},\ }\href@noop {} {\bibfield  {journal} {\bibinfo  {journal} {J.
  Chem. Phys.}\ }\textbf {\bibinfo {volume} {144}} (\bibinfo {year}
  {2016})}\BibitemShut {NoStop}%
\bibitem [{\citenamefont {Huber}\ and\ \citenamefont
  {Heller}(1988)}]{Huber:1988}%
  \BibitemOpen
  \bibfield  {author} {\bibinfo {author} {\bibfnamefont {D.}~\bibnamefont
  {Huber}}\ and\ \bibinfo {author} {\bibfnamefont {E.~J.}\ \bibnamefont
  {Heller}},\ }\href@noop {} {\bibfield  {journal} {\bibinfo  {journal} {J.
  Chem. Phys.}\ }\textbf {\bibinfo {volume} {89}},\ \bibinfo {pages} {4752}
  (\bibinfo {year} {1988})}\BibitemShut {NoStop}%
\bibitem [{\citenamefont {Mart{\'\i}nez}, \citenamefont {Ben-Nun},\ and\
  \citenamefont {Levine}(1996)}]{doi:10.1021/jp953105a}%
  \BibitemOpen
  \bibfield  {author} {\bibinfo {author} {\bibfnamefont {T.~J.}\ \bibnamefont
  {Mart{\'\i}nez}}, \bibinfo {author} {\bibfnamefont {M.}~\bibnamefont
  {Ben-Nun}}, \ and\ \bibinfo {author} {\bibfnamefont {R.~D.}\ \bibnamefont
  {Levine}},\ }\href@noop {} {\bibfield  {journal} {\bibinfo  {journal} {J.
  Phys. Chem.}\ }\textbf {\bibinfo {volume} {100}},\ \bibinfo {pages} {7884}
  (\bibinfo {year} {1996})}\BibitemShut {NoStop}%
\bibitem [{\citenamefont {Shalashilin}\ and\ \citenamefont
  {Child}(2000)}]{Shalashilin:2000}%
  \BibitemOpen
  \bibfield  {author} {\bibinfo {author} {\bibfnamefont {D.~V.}\ \bibnamefont
  {Shalashilin}}\ and\ \bibinfo {author} {\bibfnamefont {M.~S.}\ \bibnamefont
  {Child}},\ }\href@noop {} {\bibfield  {journal} {\bibinfo  {journal} {J.
  Chem. Phys.}\ }\textbf {\bibinfo {volume} {113}},\ \bibinfo {pages} {10028}
  (\bibinfo {year} {2000})}\BibitemShut {NoStop}%
\bibitem [{\citenamefont {Mauritz~Andersson}(2001)}]{Mauritz:2001}%
  \BibitemOpen
  \bibfield  {author} {\bibinfo {author} {\bibfnamefont {L.}~\bibnamefont
  {Mauritz~Andersson}},\ }\href@noop {} {\bibfield  {journal} {\bibinfo
  {journal} {The Journal of Chemical Physics}\ }\textbf {\bibinfo {volume}
  {115}},\ \bibinfo {pages} {1158} (\bibinfo {year} {2001})}\BibitemShut
  {NoStop}%
\bibitem [{\citenamefont {Wu}\ and\ \citenamefont {Batista}(2003)}]{Wu:2003}%
  \BibitemOpen
  \bibfield  {author} {\bibinfo {author} {\bibfnamefont {Y.}~\bibnamefont
  {Wu}}\ and\ \bibinfo {author} {\bibfnamefont {V.~S.}\ \bibnamefont
  {Batista}},\ }\href@noop {} {\bibfield  {journal} {\bibinfo  {journal} {The
  Journal of Chemical Physics}\ }\textbf {\bibinfo {volume} {118}},\ \bibinfo
  {pages} {6720} (\bibinfo {year} {2003})}\BibitemShut {NoStop}%
\bibitem [{\citenamefont {Koch}\ and\ \citenamefont
  {Frankcombe}(2013)}]{PhysRevLett.110.263202}%
  \BibitemOpen
  \bibfield  {author} {\bibinfo {author} {\bibfnamefont {W.}~\bibnamefont
  {Koch}}\ and\ \bibinfo {author} {\bibfnamefont {T.~J.}\ \bibnamefont
  {Frankcombe}},\ }\href@noop {} {\bibfield  {journal} {\bibinfo  {journal}
  {Phys. Rev. Lett.}\ }\textbf {\bibinfo {volume} {110}},\ \bibinfo {pages}
  {263202} (\bibinfo {year} {2013})}\BibitemShut {NoStop}%
\bibitem [{\citenamefont {Saller}\ and\ \citenamefont
  {Habershon}(2015)}]{doi:10.1021/ct500657f}%
  \BibitemOpen
  \bibfield  {author} {\bibinfo {author} {\bibfnamefont {M.~A.~C.}\
  \bibnamefont {Saller}}\ and\ \bibinfo {author} {\bibfnamefont
  {S.}~\bibnamefont {Habershon}},\ }\href@noop {} {\bibfield  {journal}
  {\bibinfo  {journal} {J. Chem. Theory Comput.}\ }\textbf {\bibinfo {volume}
  {11}},\ \bibinfo {pages} {8} (\bibinfo {year} {2015})}\BibitemShut {NoStop}%
\bibitem [{\citenamefont {Heller}(1975)}]{Heller:1975}%
  \BibitemOpen
  \bibfield  {author} {\bibinfo {author} {\bibfnamefont {E.~J.}\ \bibnamefont
  {Heller}},\ }\href@noop {} {\bibfield  {journal} {\bibinfo  {journal} {J.
  Chem. Phys.}\ }\textbf {\bibinfo {volume} {62}},\ \bibinfo {pages} {1544}
  (\bibinfo {year} {1975})}\BibitemShut {NoStop}%
\bibitem [{\citenamefont {Mart{\'\i}nez}\ and\ \citenamefont
  {Levine}(1997)}]{A605958I}%
  \BibitemOpen
  \bibfield  {author} {\bibinfo {author} {\bibfnamefont {T.~J.}\ \bibnamefont
  {Mart{\'\i}nez}}\ and\ \bibinfo {author} {\bibfnamefont {R.~D.}\ \bibnamefont
  {Levine}},\ }\href@noop {} {\bibfield  {journal} {\bibinfo  {journal} {J.
  Chem. Soc. Faraday T.}\ }\textbf {\bibinfo {volume} {93}},\ \bibinfo {pages}
  {941} (\bibinfo {year} {1997})}\BibitemShut {NoStop}%
\bibitem [{\citenamefont {Mart{\'\i}nez}, \citenamefont {Ben-Nun},\ and\
  \citenamefont {Levine}(1997)}]{doi:10.1021/jp970842t}%
  \BibitemOpen
  \bibfield  {author} {\bibinfo {author} {\bibfnamefont {T.~J.}\ \bibnamefont
  {Mart{\'\i}nez}}, \bibinfo {author} {\bibfnamefont {M.}~\bibnamefont
  {Ben-Nun}}, \ and\ \bibinfo {author} {\bibfnamefont {R.~D.}\ \bibnamefont
  {Levine}},\ }\href@noop {} {\bibfield  {journal} {\bibinfo  {journal} {J.
  Phys. Chem. A}\ }\textbf {\bibinfo {volume} {101}},\ \bibinfo {pages} {6389}
  (\bibinfo {year} {1997})}\BibitemShut {NoStop}%
\bibitem [{\citenamefont {Worth}\ and\ \citenamefont
  {Burghardt}(2003)}]{Worth2003502}%
  \BibitemOpen
  \bibfield  {author} {\bibinfo {author} {\bibfnamefont {G.~A.}\ \bibnamefont
  {Worth}}\ and\ \bibinfo {author} {\bibfnamefont {I.}~\bibnamefont
  {Burghardt}},\ }\href@noop {} {\bibfield  {journal} {\bibinfo  {journal}
  {Chem. Phys. Lett.}\ }\textbf {\bibinfo {volume} {368}},\ \bibinfo {pages}
  {502 } (\bibinfo {year} {2003})}\BibitemShut {NoStop}%
\bibitem [{\citenamefont {Fernandez-Alberti}\ \emph {et~al.}(2016)\citenamefont
  {Fernandez-Alberti}, \citenamefont {Makhov}, \citenamefont {Tretiak},\ and\
  \citenamefont {Shalashilin}}]{C5CP07332D}%
  \BibitemOpen
  \bibfield  {author} {\bibinfo {author} {\bibfnamefont {S.}~\bibnamefont
  {Fernandez-Alberti}}, \bibinfo {author} {\bibfnamefont {D.~V.}\ \bibnamefont
  {Makhov}}, \bibinfo {author} {\bibfnamefont {S.}~\bibnamefont {Tretiak}}, \
  and\ \bibinfo {author} {\bibfnamefont {D.~V.}\ \bibnamefont {Shalashilin}},\
  }\href@noop {} {\bibfield  {journal} {\bibinfo  {journal} {Phys. Chem. Chem.
  Phys.}\ }\textbf {\bibinfo {volume} {18}},\ \bibinfo {pages} {10028}
  (\bibinfo {year} {2016})}\BibitemShut {NoStop}%
\bibitem [{\citenamefont {Boyd}(2001)}]{boyd2001chebyshev}%
  \BibitemOpen
  \bibfield  {author} {\bibinfo {author} {\bibfnamefont {J.}~\bibnamefont
  {Boyd}},\ }\href@noop {} {\emph {\bibinfo {title} {Chebyshev and Fourier
  Spectral Methods: Second Revised Edition}}},\ Dover Books on Mathematics\
  (\bibinfo  {publisher} {Dover Publications},\ \bibinfo {year}
  {2001})\BibitemShut {NoStop}%
\bibitem [{\citenamefont {Lill}, \citenamefont {Parker},\ and\ \citenamefont
  {Light}(1982)}]{Lill1982483}%
  \BibitemOpen
  \bibfield  {author} {\bibinfo {author} {\bibfnamefont {J.}~\bibnamefont
  {Lill}}, \bibinfo {author} {\bibfnamefont {G.}~\bibnamefont {Parker}}, \ and\
  \bibinfo {author} {\bibfnamefont {J.}~\bibnamefont {Light}},\ }\href@noop {}
  {\bibfield  {journal} {\bibinfo  {journal} {Chem. Phys. Lett.}\ }\textbf
  {\bibinfo {volume} {89}},\ \bibinfo {pages} {483 } (\bibinfo {year}
  {1982})}\BibitemShut {NoStop}%
\bibitem [{\citenamefont {Kosloff}\ and\ \citenamefont
  {Kosloff}(1983)}]{Kosloff198335}%
  \BibitemOpen
  \bibfield  {author} {\bibinfo {author} {\bibfnamefont {D.}~\bibnamefont
  {Kosloff}}\ and\ \bibinfo {author} {\bibfnamefont {R.}~\bibnamefont
  {Kosloff}},\ }\href@noop {} {\bibfield  {journal} {\bibinfo  {journal} {J.
  Comput. Phys.}\ }\textbf {\bibinfo {volume} {52}},\ \bibinfo {pages} {35 }
  (\bibinfo {year} {1983})}\BibitemShut {NoStop}%
\bibitem [{\citenamefont {Yang}\ and\ \citenamefont {Peet}(1988)}]{Yang198898}%
  \BibitemOpen
  \bibfield  {author} {\bibinfo {author} {\bibfnamefont {W.}~\bibnamefont
  {Yang}}\ and\ \bibinfo {author} {\bibfnamefont {A.~C.}\ \bibnamefont
  {Peet}},\ }\href@noop {} {\bibfield  {journal} {\bibinfo  {journal} {Chem.
  Phys. Lett.}\ }\textbf {\bibinfo {volume} {153}},\ \bibinfo {pages} {98 }
  (\bibinfo {year} {1988})}\BibitemShut {NoStop}%
\bibitem [{\citenamefont {Kosloff}(1988)}]{doi:10.1021/j100319a003}%
  \BibitemOpen
  \bibfield  {author} {\bibinfo {author} {\bibfnamefont {R.}~\bibnamefont
  {Kosloff}},\ }\href@noop {} {\bibfield  {journal} {\bibinfo  {journal} {J.
  Phys. Chem.}\ }\textbf {\bibinfo {volume} {92}},\ \bibinfo {pages} {2087}
  (\bibinfo {year} {1988})}\BibitemShut {NoStop}%
\bibitem [{\citenamefont {Peet}\ and\ \citenamefont
  {Yang}(1989{\natexlab{a}})}]{Peet:1989a}%
  \BibitemOpen
  \bibfield  {author} {\bibinfo {author} {\bibfnamefont {A.~C.}\ \bibnamefont
  {Peet}}\ and\ \bibinfo {author} {\bibfnamefont {W.}~\bibnamefont {Yang}},\
  }\href@noop {} {\bibfield  {journal} {\bibinfo  {journal} {J. Chem. Phys.}\
  }\textbf {\bibinfo {volume} {91}},\ \bibinfo {pages} {6598} (\bibinfo {year}
  {1989}{\natexlab{a}})}\BibitemShut {NoStop}%
\bibitem [{\citenamefont {Peet}\ and\ \citenamefont
  {Yang}(1989{\natexlab{b}})}]{Peet:1989}%
  \BibitemOpen
  \bibfield  {author} {\bibinfo {author} {\bibfnamefont {A.~C.}\ \bibnamefont
  {Peet}}\ and\ \bibinfo {author} {\bibfnamefont {W.}~\bibnamefont {Yang}},\
  }\href@noop {} {\bibfield  {journal} {\bibinfo  {journal} {J. Chem. Phys.}\
  }\textbf {\bibinfo {volume} {90}},\ \bibinfo {pages} {1746} (\bibinfo {year}
  {1989}{\natexlab{b}})}\BibitemShut {NoStop}%
\bibitem [{\citenamefont {Yang}\ and\ \citenamefont {Peet}(1990)}]{Peet:1990a}%
  \BibitemOpen
  \bibfield  {author} {\bibinfo {author} {\bibfnamefont {W.}~\bibnamefont
  {Yang}}\ and\ \bibinfo {author} {\bibfnamefont {A.~C.}\ \bibnamefont
  {Peet}},\ }\href@noop {} {\bibfield  {journal} {\bibinfo  {journal} {J. Chem.
  Phys.}\ }\textbf {\bibinfo {volume} {92}},\ \bibinfo {pages} {522} (\bibinfo
  {year} {1990})}\BibitemShut {NoStop}%
\bibitem [{\citenamefont {Sielk}\ \emph {et~al.}(2009)\citenamefont {Sielk},
  \citenamefont {von Horsten}, \citenamefont {Kruger}, \citenamefont
  {Schneider},\ and\ \citenamefont {Hartke}}]{B814315C}%
  \BibitemOpen
  \bibfield  {author} {\bibinfo {author} {\bibfnamefont {J.}~\bibnamefont
  {Sielk}}, \bibinfo {author} {\bibfnamefont {H.~F.}\ \bibnamefont {von
  Horsten}}, \bibinfo {author} {\bibfnamefont {F.}~\bibnamefont {Kruger}},
  \bibinfo {author} {\bibfnamefont {R.}~\bibnamefont {Schneider}}, \ and\
  \bibinfo {author} {\bibfnamefont {B.}~\bibnamefont {Hartke}},\ }\href@noop {}
  {\bibfield  {journal} {\bibinfo  {journal} {Phys. Chem. Chem. Phys.}\
  }\textbf {\bibinfo {volume} {11}},\ \bibinfo {pages} {463} (\bibinfo {year}
  {2009})}\BibitemShut {NoStop}%
\bibitem [{\citenamefont {Yang}, \citenamefont {Peet},\ and\ \citenamefont
  {Miller}(1989)}]{Yang:1989b}%
  \BibitemOpen
  \bibfield  {author} {\bibinfo {author} {\bibfnamefont {W.}~\bibnamefont
  {Yang}}, \bibinfo {author} {\bibfnamefont {A.~C.}\ \bibnamefont {Peet}}, \
  and\ \bibinfo {author} {\bibfnamefont {W.~H.}\ \bibnamefont {Miller}},\
  }\href@noop {} {\bibfield  {journal} {\bibinfo  {journal} {J. Chem. Phys.}\
  }\textbf {\bibinfo {volume} {91}},\ \bibinfo {pages} {7537} (\bibinfo {year}
  {1989})}\BibitemShut {NoStop}%
\bibitem [{\citenamefont {Dehghan}\ and\ \citenamefont
  {Shokri}(2007)}]{Dehghan2007136}%
  \BibitemOpen
  \bibfield  {author} {\bibinfo {author} {\bibfnamefont {M.}~\bibnamefont
  {Dehghan}}\ and\ \bibinfo {author} {\bibfnamefont {A.}~\bibnamefont
  {Shokri}},\ }\href@noop {} {\bibfield  {journal} {\bibinfo  {journal}
  {Comput. Math. Appl.}\ }\textbf {\bibinfo {volume} {54}},\ \bibinfo {pages}
  {136 } (\bibinfo {year} {2007})}\BibitemShut {NoStop}%
\bibitem [{\citenamefont {Orszag}(1969)}]{orszag:1969}%
  \BibitemOpen
  \bibfield  {author} {\bibinfo {author} {\bibfnamefont {S.~A.}\ \bibnamefont
  {Orszag}},\ }\href@noop {} {\bibfield  {journal} {\bibinfo  {journal} {Phys.
  Fluids}\ }\textbf {\bibinfo {volume} {12}},\ \bibinfo {pages} {250} (\bibinfo
  {year} {1969})}\BibitemShut {NoStop}%
\bibitem [{\citenamefont {Gottlieb}\ and\ \citenamefont
  {Orszag}(1977)}]{gottlieb1977numerical}%
  \BibitemOpen
  \bibfield  {author} {\bibinfo {author} {\bibfnamefont {D.}~\bibnamefont
  {Gottlieb}}\ and\ \bibinfo {author} {\bibfnamefont {S.}~\bibnamefont
  {Orszag}},\ }\href@noop {} {\emph {\bibinfo {title} {Numerical Analysis of
  Spectral Methods: Theory and Applications}}}\ (\bibinfo  {publisher} {SIAM},\
  \bibinfo {year} {1977})\BibitemShut {NoStop}%
\bibitem [{\citenamefont {Furnaro}(1992)}]{furnaro:1992}%
  \BibitemOpen
  \bibfield  {author} {\bibinfo {author} {\bibfnamefont {D.}~\bibnamefont
  {Furnaro}},\ }in\ \href@noop {} {\emph {\bibinfo {booktitle} {Polynomial
  Approximation of Differential Equations}}},\ \bibinfo {series} {Lecture Notes
  in Physics}, Vol.~\bibinfo {volume} {8}\ (\bibinfo  {publisher}
  {Springer-Verlag Heidelberg},\ \bibinfo {year} {1992})\BibitemShut {NoStop}%
\bibitem [{\citenamefont {Fornberg}(1998)}]{fornberg1998practical}%
  \BibitemOpen
  \bibfield  {author} {\bibinfo {author} {\bibfnamefont {B.}~\bibnamefont
  {Fornberg}},\ }\href@noop {} {\emph {\bibinfo {title} {A Practical Guide to
  Pseudospectral Methods}}}\ (\bibinfo  {publisher} {Cambridge University
  Press},\ \bibinfo {year} {1998})\BibitemShut {NoStop}%
\bibitem [{\citenamefont {Canuto}\ \emph {et~al.}(2006)\citenamefont {Canuto},
  \citenamefont {Hussaini}, \citenamefont {Quarteroni},\ and\ \citenamefont
  {Zang}}]{Canuto:2006}%
  \BibitemOpen
  \bibfield  {author} {\bibinfo {author} {\bibfnamefont {C.}~\bibnamefont
  {Canuto}}, \bibinfo {author} {\bibfnamefont {M.~Y.}\ \bibnamefont
  {Hussaini}}, \bibinfo {author} {\bibfnamefont {A.}~\bibnamefont
  {Quarteroni}}, \ and\ \bibinfo {author} {\bibfnamefont {T.}~\bibnamefont
  {Zang}},\ }\href@noop {} {\emph {\bibinfo {title} {Spectral Methods:
  Fundamentals in Single Domains}}}\ (\bibinfo  {publisher} {Springer-Verlag
  Berlin Heidelberg},\ \bibinfo {year} {2006})\BibitemShut {NoStop}%
\bibitem [{\citenamefont {Hesthaven}, \citenamefont {Gottlieb},\ and\
  \citenamefont {Gottlieb}(2007)}]{hesthaven2007spectral}%
  \BibitemOpen
  \bibfield  {author} {\bibinfo {author} {\bibfnamefont {J.~S.}\ \bibnamefont
  {Hesthaven}}, \bibinfo {author} {\bibfnamefont {S.}~\bibnamefont {Gottlieb}},
  \ and\ \bibinfo {author} {\bibfnamefont {D.}~\bibnamefont {Gottlieb}},\
  }\href@noop {} {\emph {\bibinfo {title} {Spectral methods for time-dependent
  problems}}}\ (\bibinfo  {publisher} {Cambridge University Press},\ \bibinfo
  {year} {2007})\BibitemShut {NoStop}%
\bibitem [{\citenamefont {Tannor}(2007)}]{tannor2007introduction}%
  \BibitemOpen
  \bibfield  {author} {\bibinfo {author} {\bibfnamefont {D.}~\bibnamefont
  {Tannor}},\ }\href@noop {} {\emph {\bibinfo {title} {Introduction to Quantum
  Mechanics: A Time-dependent Perspective}}}\ (\bibinfo  {publisher}
  {University Science Books},\ \bibinfo {year} {2007})\BibitemShut {NoStop}%
\bibitem [{\citenamefont {Billing}(1983)}]{BILLING1983535}%
  \BibitemOpen
  \bibfield  {author} {\bibinfo {author} {\bibfnamefont {G.~D.}\ \bibnamefont
  {Billing}},\ }\href@noop {} {\bibfield  {journal} {\bibinfo  {journal} {Chem.
  Phys. Lett.}\ }\textbf {\bibinfo {volume} {100}},\ \bibinfo {pages} {535 }
  (\bibinfo {year} {1983})}\BibitemShut {NoStop}%
\bibitem [{\citenamefont {Tully}(1990)}]{tully:1061}%
  \BibitemOpen
  \bibfield  {author} {\bibinfo {author} {\bibfnamefont {J.~C.}\ \bibnamefont
  {Tully}},\ }\href@noop {} {\bibfield  {journal} {\bibinfo  {journal} {J.
  Chem. Phys.}\ }\textbf {\bibinfo {volume} {93}},\ \bibinfo {pages} {1061}
  (\bibinfo {year} {1990})}\BibitemShut {NoStop}%
\bibitem [{\citenamefont {C.~Tully}(1998)}]{A801824C}%
  \BibitemOpen
  \bibfield  {author} {\bibinfo {author} {\bibfnamefont {J.}~\bibnamefont
  {C.~Tully}},\ }\href@noop {} {\bibfield  {journal} {\bibinfo  {journal}
  {Faraday Discuss.}\ }\textbf {\bibinfo {volume} {110}},\ \bibinfo {pages}
  {407} (\bibinfo {year} {1998})}\BibitemShut {NoStop}%
\bibitem [{\citenamefont {Hack}\ and\ \citenamefont
  {Truhlar}(2000)}]{Hack:2000}%
  \BibitemOpen
  \bibfield  {author} {\bibinfo {author} {\bibfnamefont {M.~D.}\ \bibnamefont
  {Hack}}\ and\ \bibinfo {author} {\bibfnamefont {D.~G.}\ \bibnamefont
  {Truhlar}},\ }\href@noop {} {\bibfield  {journal} {\bibinfo  {journal} {J.
  Phys. Chem. A}\ }\textbf {\bibinfo {volume} {104}},\ \bibinfo {pages} {7917}
  (\bibinfo {year} {2000})}\BibitemShut {NoStop}%
\bibitem [{\citenamefont
  {Shalashilin}(2009)}]{:/content/aip/journal/jcp/130/24/10.1063/1.3153302}%
  \BibitemOpen
  \bibfield  {author} {\bibinfo {author} {\bibfnamefont {D.~V.}\ \bibnamefont
  {Shalashilin}},\ }\href@noop {} {\bibfield  {journal} {\bibinfo  {journal}
  {J. Chem. Phys.}\ }\textbf {\bibinfo {volume} {130}},\ \bibinfo {pages}
  {244101} (\bibinfo {year} {2009})}\BibitemShut {NoStop}%
\bibitem [{\citenamefont {Subotnik}(2010)}]{Subotnik:2010fk}%
  \BibitemOpen
  \bibfield  {author} {\bibinfo {author} {\bibfnamefont {J.~E.}\ \bibnamefont
  {Subotnik}},\ }\href@noop {} {\bibfield  {journal} {\bibinfo  {journal} {J.
  Chem. Phys.}\ }\textbf {\bibinfo {volume} {132}},\ \bibinfo {pages} {134112}
  (\bibinfo {year} {2010})}\BibitemShut {NoStop}%
\bibitem [{\citenamefont {Mart{\'\i}nez}, \citenamefont {Ben-Nun},\ and\
  \citenamefont {Ashkenazi}(1996)}]{Martinez:1996}%
  \BibitemOpen
  \bibfield  {author} {\bibinfo {author} {\bibfnamefont {T.~J.}\ \bibnamefont
  {Mart{\'\i}nez}}, \bibinfo {author} {\bibfnamefont {M.}~\bibnamefont
  {Ben-Nun}}, \ and\ \bibinfo {author} {\bibfnamefont {G.}~\bibnamefont
  {Ashkenazi}},\ }\href@noop {} {\bibfield  {journal} {\bibinfo  {journal} {J.
  Chem. Phys.}\ }\textbf {\bibinfo {volume} {104}},\ \bibinfo {pages} {2847}
  (\bibinfo {year} {1996})}\BibitemShut {NoStop}%
\bibitem [{\citenamefont {Mart{\'\i}nez}(2006)}]{doi:10.1021/ar040202q}%
  \BibitemOpen
  \bibfield  {author} {\bibinfo {author} {\bibfnamefont {T.~J.}\ \bibnamefont
  {Mart{\'\i}nez}},\ }\href@noop {} {\bibfield  {journal} {\bibinfo  {journal}
  {Acc. Chem. Res.}\ }\textbf {\bibinfo {volume} {39}},\ \bibinfo {pages} {119}
  (\bibinfo {year} {2006})}\BibitemShut {NoStop}%
\bibitem [{\citenamefont {Yang}\ \emph {et~al.}(2009)\citenamefont {Yang},
  \citenamefont {Coe}, \citenamefont {Kaduk},\ and\ \citenamefont
  {Mart{\'\i}nez}}]{:/content/aip/journal/jcp/130/13/10.1063/1.3103930}%
  \BibitemOpen
  \bibfield  {author} {\bibinfo {author} {\bibfnamefont {S.}~\bibnamefont
  {Yang}}, \bibinfo {author} {\bibfnamefont {J.~D.}\ \bibnamefont {Coe}},
  \bibinfo {author} {\bibfnamefont {B.}~\bibnamefont {Kaduk}}, \ and\ \bibinfo
  {author} {\bibfnamefont {T.~J.}\ \bibnamefont {Mart{\'\i}nez}},\ }\href@noop
  {} {\bibfield  {journal} {\bibinfo  {journal} {J. Chem. Phys.}\ }\textbf
  {\bibinfo {volume} {130}},\ \bibinfo {pages} {134113} (\bibinfo {year}
  {2009})}\BibitemShut {NoStop}%
\bibitem [{\citenamefont {Subotnik}\ and\ \citenamefont
  {Shenvi}(2011{\natexlab{a}})}]{Subotnik:2011fk}%
  \BibitemOpen
  \bibfield  {author} {\bibinfo {author} {\bibfnamefont {J.~E.}\ \bibnamefont
  {Subotnik}}\ and\ \bibinfo {author} {\bibfnamefont {N.}~\bibnamefont
  {Shenvi}},\ }\href@noop {} {\bibfield  {journal} {\bibinfo  {journal} {J.
  Chem. Phys.}\ }\textbf {\bibinfo {volume} {134}},\ \bibinfo {pages} {024105}
  (\bibinfo {year} {2011}{\natexlab{a}})}\BibitemShut {NoStop}%
\bibitem [{\citenamefont {Wyatt}, \citenamefont {Lopreore},\ and\ \citenamefont
  {Parlant}(2001)}]{wyatt:5113}%
  \BibitemOpen
  \bibfield  {author} {\bibinfo {author} {\bibfnamefont {R.~E.}\ \bibnamefont
  {Wyatt}}, \bibinfo {author} {\bibfnamefont {C.~L.}\ \bibnamefont {Lopreore}},
  \ and\ \bibinfo {author} {\bibfnamefont {G.}~\bibnamefont {Parlant}},\
  }\href@noop {} {\bibfield  {journal} {\bibinfo  {journal} {J. Chem. Phys.}\
  }\textbf {\bibinfo {volume} {114}},\ \bibinfo {pages} {5113} (\bibinfo {year}
  {2001})}\BibitemShut {NoStop}%
\bibitem [{\citenamefont {Rassolov}\ and\ \citenamefont
  {Garashchuk}(2005)}]{PhysRevA.71.032511}%
  \BibitemOpen
  \bibfield  {author} {\bibinfo {author} {\bibfnamefont {V.~A.}\ \bibnamefont
  {Rassolov}}\ and\ \bibinfo {author} {\bibfnamefont {S.}~\bibnamefont
  {Garashchuk}},\ }\href@noop {} {\bibfield  {journal} {\bibinfo  {journal}
  {Phys. Rev. A}\ }\textbf {\bibinfo {volume} {71}},\ \bibinfo {pages} {032511}
  (\bibinfo {year} {2005})}\BibitemShut {NoStop}%
\bibitem [{\citenamefont {Curchod}, \citenamefont {Tavernelli},\ and\
  \citenamefont {Rothlisberger}(2011)}]{C0CP02175J}%
  \BibitemOpen
  \bibfield  {author} {\bibinfo {author} {\bibfnamefont {B.~F.~E.}\
  \bibnamefont {Curchod}}, \bibinfo {author} {\bibfnamefont {I.}~\bibnamefont
  {Tavernelli}}, \ and\ \bibinfo {author} {\bibfnamefont {U.}~\bibnamefont
  {Rothlisberger}},\ }\href@noop {} {\bibfield  {journal} {\bibinfo  {journal}
  {Phys. Chem. Chem. Phys.}\ }\textbf {\bibinfo {volume} {13}},\ \bibinfo
  {pages} {3231} (\bibinfo {year} {2011})}\BibitemShut {NoStop}%
\bibitem [{\citenamefont {Zamstein}\ and\ \citenamefont
  {Tannor}(2012{\natexlab{a}})}]{zamstein:22A517}%
  \BibitemOpen
  \bibfield  {author} {\bibinfo {author} {\bibfnamefont {N.}~\bibnamefont
  {Zamstein}}\ and\ \bibinfo {author} {\bibfnamefont {D.~J.}\ \bibnamefont
  {Tannor}},\ }\href@noop {} {\bibfield  {journal} {\bibinfo  {journal} {J.
  Chem. Phys.}\ }\textbf {\bibinfo {volume} {137}},\ \bibinfo {pages} {22A517}
  (\bibinfo {year} {2012}{\natexlab{a}})}\BibitemShut {NoStop}%
\bibitem [{\citenamefont {Zamstein}\ and\ \citenamefont
  {Tannor}(2012{\natexlab{b}})}]{zamstein:22A518}%
  \BibitemOpen
  \bibfield  {author} {\bibinfo {author} {\bibfnamefont {N.}~\bibnamefont
  {Zamstein}}\ and\ \bibinfo {author} {\bibfnamefont {D.~J.}\ \bibnamefont
  {Tannor}},\ }\href@noop {} {\bibfield  {journal} {\bibinfo  {journal} {J.
  Chem. Phys.}\ }\textbf {\bibinfo {volume} {137}},\ \bibinfo {pages} {22A518}
  (\bibinfo {year} {2012}{\natexlab{b}})}\BibitemShut {NoStop}%
\bibitem [{\citenamefont {Curchod}\ and\ \citenamefont
  {Tavernelli}(2013)}]{curchod:184112}%
  \BibitemOpen
  \bibfield  {author} {\bibinfo {author} {\bibfnamefont {B.~F.~E.}\
  \bibnamefont {Curchod}}\ and\ \bibinfo {author} {\bibfnamefont
  {I.}~\bibnamefont {Tavernelli}},\ }\href@noop {} {\bibfield  {journal}
  {\bibinfo  {journal} {J. Chem. Phys.}\ }\textbf {\bibinfo {volume} {138}},\
  \bibinfo {pages} {184112} (\bibinfo {year} {2013})}\BibitemShut {NoStop}%
\bibitem [{\citenamefont {Bittner}\ and\ \citenamefont
  {Rossky}(1995)}]{Bittner:1995}%
  \BibitemOpen
  \bibfield  {author} {\bibinfo {author} {\bibfnamefont {E.~R.}\ \bibnamefont
  {Bittner}}\ and\ \bibinfo {author} {\bibfnamefont {P.~J.}\ \bibnamefont
  {Rossky}},\ }\href@noop {} {\bibfield  {journal} {\bibinfo  {journal} {J.
  Chem. Phys.}\ }\textbf {\bibinfo {volume} {103}},\ \bibinfo {pages} {8130}
  (\bibinfo {year} {1995})}\BibitemShut {NoStop}%
\bibitem [{\citenamefont {Granucci}, \citenamefont {Persico},\ and\
  \citenamefont
  {Zoccante}(2010)}]{:/content/aip/journal/jcp/133/13/10.1063/1.3489004}%
  \BibitemOpen
  \bibfield  {author} {\bibinfo {author} {\bibfnamefont {G.}~\bibnamefont
  {Granucci}}, \bibinfo {author} {\bibfnamefont {M.}~\bibnamefont {Persico}}, \
  and\ \bibinfo {author} {\bibfnamefont {A.}~\bibnamefont {Zoccante}},\
  }\href@noop {} {\bibfield  {journal} {\bibinfo  {journal} {J. Chem. Phys.}\
  }\textbf {\bibinfo {volume} {133}} (\bibinfo {year} {2010})}\BibitemShut
  {NoStop}%
\bibitem [{\citenamefont {Landry}\ and\ \citenamefont
  {Subotnik}(2011)}]{landry:191101}%
  \BibitemOpen
  \bibfield  {author} {\bibinfo {author} {\bibfnamefont {B.~R.}\ \bibnamefont
  {Landry}}\ and\ \bibinfo {author} {\bibfnamefont {J.~E.}\ \bibnamefont
  {Subotnik}},\ }\href@noop {} {\bibfield  {journal} {\bibinfo  {journal} {J.
  Chem. Phys.}\ }\textbf {\bibinfo {volume} {135}},\ \bibinfo {pages} {191101}
  (\bibinfo {year} {2011})}\BibitemShut {NoStop}%
\bibitem [{\citenamefont {Subotnik}(2011)}]{doi:10.1021/jp206557h}%
  \BibitemOpen
  \bibfield  {author} {\bibinfo {author} {\bibfnamefont {J.~E.}\ \bibnamefont
  {Subotnik}},\ }\href@noop {} {\bibfield  {journal} {\bibinfo  {journal} {J.
  Phys. Chem. A}\ }\textbf {\bibinfo {volume} {115}},\ \bibinfo {pages} {12083}
  (\bibinfo {year} {2011})}\BibitemShut {NoStop}%
\bibitem [{\citenamefont {Shenvi}, \citenamefont {Subotnik},\ and\
  \citenamefont
  {Yang}(2011{\natexlab{a}})}]{:/content/aip/journal/jcp/135/2/10.1063/1.3603447}%
  \BibitemOpen
  \bibfield  {author} {\bibinfo {author} {\bibfnamefont {N.}~\bibnamefont
  {Shenvi}}, \bibinfo {author} {\bibfnamefont {J.~E.}\ \bibnamefont
  {Subotnik}}, \ and\ \bibinfo {author} {\bibfnamefont {W.}~\bibnamefont
  {Yang}},\ }\href@noop {} {\bibfield  {journal} {\bibinfo  {journal} {J. Chem.
  Phys.}\ }\textbf {\bibinfo {volume} {135}},\ \bibinfo {pages} {024101}
  (\bibinfo {year} {2011}{\natexlab{a}})}\BibitemShut {NoStop}%
\bibitem [{\citenamefont {Shenvi}, \citenamefont {Subotnik},\ and\
  \citenamefont {Yang}(2011{\natexlab{b}})}]{Shenvi:2011uq}%
  \BibitemOpen
  \bibfield  {author} {\bibinfo {author} {\bibfnamefont {N.}~\bibnamefont
  {Shenvi}}, \bibinfo {author} {\bibfnamefont {J.~E.}\ \bibnamefont
  {Subotnik}}, \ and\ \bibinfo {author} {\bibfnamefont {W.}~\bibnamefont
  {Yang}},\ }\href@noop {} {\bibfield  {journal} {\bibinfo  {journal} {J. Chem.
  Phys.}\ }\textbf {\bibinfo {volume} {134}},\ \bibinfo {pages} {144102}
  (\bibinfo {year} {2011}{\natexlab{b}})}\BibitemShut {NoStop}%
\bibitem [{\citenamefont {Subotnik}\ and\ \citenamefont
  {Shenvi}(2011{\natexlab{b}})}]{Subotnik:2011uq}%
  \BibitemOpen
  \bibfield  {author} {\bibinfo {author} {\bibfnamefont {J.~E.}\ \bibnamefont
  {Subotnik}}\ and\ \bibinfo {author} {\bibfnamefont {N.}~\bibnamefont
  {Shenvi}},\ }\href@noop {} {\bibfield  {journal} {\bibinfo  {journal} {J.
  Chem. Phys.}\ }\textbf {\bibinfo {volume} {134}},\ \bibinfo {pages} {244114}
  (\bibinfo {year} {2011}{\natexlab{b}})}\BibitemShut {NoStop}%
\bibitem [{\citenamefont {Landry}\ and\ \citenamefont
  {Subotnik}(2012)}]{landry:22A513}%
  \BibitemOpen
  \bibfield  {author} {\bibinfo {author} {\bibfnamefont {B.~R.}\ \bibnamefont
  {Landry}}\ and\ \bibinfo {author} {\bibfnamefont {J.~E.}\ \bibnamefont
  {Subotnik}},\ }\href@noop {} {\bibfield  {journal} {\bibinfo  {journal} {J.
  Chem. Phys.}\ }\textbf {\bibinfo {volume} {137}},\ \bibinfo {pages} {22A513}
  (\bibinfo {year} {2012})}\BibitemShut {NoStop}%
\bibitem [{\citenamefont {Coronado}, \citenamefont {Xing},\ and\ \citenamefont
  {Miller}(2001)}]{Coronado2001521}%
  \BibitemOpen
  \bibfield  {author} {\bibinfo {author} {\bibfnamefont {E.~A.}\ \bibnamefont
  {Coronado}}, \bibinfo {author} {\bibfnamefont {J.}~\bibnamefont {Xing}}, \
  and\ \bibinfo {author} {\bibfnamefont {W.~H.}\ \bibnamefont {Miller}},\
  }\href@noop {} {\bibfield  {journal} {\bibinfo  {journal} {Chem. Phys.
  Lett.}\ }\textbf {\bibinfo {volume} {349}},\ \bibinfo {pages} {521 }
  (\bibinfo {year} {2001})}\BibitemShut {NoStop}%
\bibitem [{\citenamefont {Bonella}\ and\ \citenamefont
  {Coker}(2003)}]{Bonella:2003}%
  \BibitemOpen
  \bibfield  {author} {\bibinfo {author} {\bibfnamefont {S.}~\bibnamefont
  {Bonella}}\ and\ \bibinfo {author} {\bibfnamefont {D.~F.}\ \bibnamefont
  {Coker}},\ }\href@noop {} {\bibfield  {journal} {\bibinfo  {journal} {J.
  Chem. Phys.}\ }\textbf {\bibinfo {volume} {118}},\ \bibinfo {pages} {4370}
  (\bibinfo {year} {2003})}\BibitemShut {NoStop}%
\bibitem [{\citenamefont {Ferretti}\ \emph {et~al.}(1996)\citenamefont
  {Ferretti}, \citenamefont {Granucci}, \citenamefont {Lami}, \citenamefont
  {Persico},\ and\ \citenamefont {Villani}}]{Ferretti:1996}%
  \BibitemOpen
  \bibfield  {author} {\bibinfo {author} {\bibfnamefont {A.}~\bibnamefont
  {Ferretti}}, \bibinfo {author} {\bibfnamefont {G.}~\bibnamefont {Granucci}},
  \bibinfo {author} {\bibfnamefont {A.}~\bibnamefont {Lami}}, \bibinfo {author}
  {\bibfnamefont {M.}~\bibnamefont {Persico}}, \ and\ \bibinfo {author}
  {\bibfnamefont {G.}~\bibnamefont {Villani}},\ }\href@noop {} {\bibfield
  {journal} {\bibinfo  {journal} {J. Chem. Phys.}\ }\textbf {\bibinfo {volume}
  {104}},\ \bibinfo {pages} {5517} (\bibinfo {year} {1996})}\BibitemShut
  {NoStop}%
\bibitem [{\citenamefont {Born}\ and\ \citenamefont
  {Huang}(1998)}]{born1998dynamical}%
  \BibitemOpen
  \bibfield  {author} {\bibinfo {author} {\bibfnamefont {M.}~\bibnamefont
  {Born}}\ and\ \bibinfo {author} {\bibfnamefont {K.}~\bibnamefont {Huang}},\
  }\href@noop {} {\emph {\bibinfo {title} {Dynamical Theory of Crystal
  Lattices}}},\ International series of monographs on physics\ (\bibinfo
  {publisher} {Clarendon Press},\ \bibinfo {year} {1998})\BibitemShut {NoStop}%
\bibitem [{\citenamefont {Finlayson}(1972)}]{finlayson1972method}%
  \BibitemOpen
  \bibfield  {author} {\bibinfo {author} {\bibfnamefont {B.}~\bibnamefont
  {Finlayson}},\ }\href@noop {} {\emph {\bibinfo {title} {The Method of
  Weighted Residuals and Variational Principles: With Application in Fluid
  Mechanics, Heat and Mass Transfer}}},\ Educational Psychology\ (\bibinfo
  {publisher} {Academic Press},\ \bibinfo {year} {1972})\BibitemShut {NoStop}%
\bibitem [{\citenamefont {Frost}(1964)}]{F64}%
  \BibitemOpen
  \bibfield  {author} {\bibinfo {author} {\bibfnamefont {A.~A.}\ \bibnamefont
  {Frost}},\ }\href@noop {} {\bibfield  {journal} {\bibinfo  {journal} {J.
  Chem. Phys.}\ }\textbf {\bibinfo {volume} {41}},\ \bibinfo {pages} {478}
  (\bibinfo {year} {1964})}\BibitemShut {NoStop}%
\bibitem [{\citenamefont {Heller}(1981)}]{Heller:1981}%
  \BibitemOpen
  \bibfield  {author} {\bibinfo {author} {\bibfnamefont {E.~J.}\ \bibnamefont
  {Heller}},\ }\href@noop {} {\bibfield  {journal} {\bibinfo  {journal} {J.
  Chem. Phys.}\ }\textbf {\bibinfo {volume} {75}},\ \bibinfo {pages} {2923}
  (\bibinfo {year} {1981})}\BibitemShut {NoStop}%
\bibitem [{\citenamefont {Hansen}(2010)}]{hansen2010discrete}%
  \BibitemOpen
  \bibfield  {author} {\bibinfo {author} {\bibfnamefont {P.}~\bibnamefont
  {Hansen}},\ }\href@noop {} {\emph {\bibinfo {title} {Discrete Inverse
  Problems: Insight and Algorithms}}},\ Fundamentals of Algorithms\ (\bibinfo
  {publisher} {SIAM},\ \bibinfo {year} {2010})\BibitemShut {NoStop}%
\bibitem [{\citenamefont {Ben-Nun}\ and\ \citenamefont
  {Mart{\'\i}nez}(1998{\natexlab{b}})}]{ben-nun:7244}%
  \BibitemOpen
  \bibfield  {author} {\bibinfo {author} {\bibfnamefont {M.}~\bibnamefont
  {Ben-Nun}}\ and\ \bibinfo {author} {\bibfnamefont {T.~J.}\ \bibnamefont
  {Mart{\'\i}nez}},\ }\href@noop {} {\bibfield  {journal} {\bibinfo  {journal}
  {J. Chem. Phys.}\ }\textbf {\bibinfo {volume} {108}},\ \bibinfo {pages}
  {7244} (\bibinfo {year} {1998}{\natexlab{b}})}\BibitemShut {NoStop}%
\bibitem [{\citenamefont {Burghardt}, \citenamefont {Nest},\ and\ \citenamefont
  {Worth}(2003)}]{Burghardt:2003}%
  \BibitemOpen
  \bibfield  {author} {\bibinfo {author} {\bibfnamefont {I.}~\bibnamefont
  {Burghardt}}, \bibinfo {author} {\bibfnamefont {M.}~\bibnamefont {Nest}}, \
  and\ \bibinfo {author} {\bibfnamefont {G.~A.}\ \bibnamefont {Worth}},\
  }\href@noop {} {\bibfield  {journal} {\bibinfo  {journal} {J. Chem. Phys.}\
  }\textbf {\bibinfo {volume} {119}},\ \bibinfo {pages} {5364} (\bibinfo {year}
  {2003})}\BibitemShut {NoStop}%
\bibitem [{\citenamefont {Colbert}\ and\ \citenamefont
  {Miller}(1992)}]{Colbert:1992}%
  \BibitemOpen
  \bibfield  {author} {\bibinfo {author} {\bibfnamefont {D.~T.}\ \bibnamefont
  {Colbert}}\ and\ \bibinfo {author} {\bibfnamefont {W.~H.}\ \bibnamefont
  {Miller}},\ }\href@noop {} {\bibfield  {journal} {\bibinfo  {journal} {J.
  Chem. Phys.}\ }\textbf {\bibinfo {volume} {96}},\ \bibinfo {pages} {1982}
  (\bibinfo {year} {1992})}\BibitemShut {NoStop}%
\bibitem [{\citenamefont {Light}\ and\ \citenamefont
  {Carrington}(2000)}]{Light:2000}%
  \BibitemOpen
  \bibfield  {author} {\bibinfo {author} {\bibfnamefont {J.~C.}\ \bibnamefont
  {Light}}\ and\ \bibinfo {author} {\bibfnamefont {T.}~\bibnamefont
  {Carrington}},\ }\href@noop {} {\bibfield  {journal} {\bibinfo  {journal}
  {Adv. Chem. Phys.}\ }\textbf {\bibinfo {volume} {114}},\ \bibinfo {pages}
  {263} (\bibinfo {year} {2000})}\BibitemShut {NoStop}%
\bibitem [{\citenamefont {Mazziotti}(1999)}]{Mazziotti1999473}%
  \BibitemOpen
  \bibfield  {author} {\bibinfo {author} {\bibfnamefont {D.~A.}\ \bibnamefont
  {Mazziotti}},\ }\href@noop {} {\bibfield  {journal} {\bibinfo  {journal}
  {Chem. Phys. Lett.}\ }\textbf {\bibinfo {volume} {299}},\ \bibinfo {pages}
  {473 } (\bibinfo {year} {1999})}\BibitemShut {NoStop}%
\bibitem [{\citenamefont {Mazziotti}(2002)}]{Mazziotti:2002SD}%
  \BibitemOpen
  \bibfield  {author} {\bibinfo {author} {\bibfnamefont {D.~A.}\ \bibnamefont
  {Mazziotti}},\ }\href@noop {} {\bibfield  {journal} {\bibinfo  {journal} {J.
  Chem. Phys.}\ }\textbf {\bibinfo {volume} {117}},\ \bibinfo {pages} {2455}
  (\bibinfo {year} {2002})}\BibitemShut {NoStop}%
\bibitem [{\citenamefont {Wigner}(1932)}]{PhysRev.40.749}%
  \BibitemOpen
  \bibfield  {author} {\bibinfo {author} {\bibfnamefont {E.}~\bibnamefont
  {Wigner}},\ }\href@noop {} {\bibfield  {journal} {\bibinfo  {journal} {Phys.
  Rev.}\ }\textbf {\bibinfo {volume} {40}},\ \bibinfo {pages} {749} (\bibinfo
  {year} {1932})}\BibitemShut {NoStop}%
\bibitem [{\citenamefont {Heller}(1976)}]{Heller:1976a}%
  \BibitemOpen
  \bibfield  {author} {\bibinfo {author} {\bibfnamefont {E.~J.}\ \bibnamefont
  {Heller}},\ }\href@noop {} {\bibfield  {journal} {\bibinfo  {journal} {J.
  Chem. Phys.}\ }\textbf {\bibinfo {volume} {65}},\ \bibinfo {pages} {1289}
  (\bibinfo {year} {1976})}\BibitemShut {NoStop}%
\bibitem [{\citenamefont {Yeganeh}\ and\ \citenamefont
  {Ratner}(2006)}]{:/content/aip/journal/jcp/124/4/10.1063/1.2162172}%
  \BibitemOpen
  \bibfield  {author} {\bibinfo {author} {\bibfnamefont {S.}~\bibnamefont
  {Yeganeh}}\ and\ \bibinfo {author} {\bibfnamefont {M.~A.}\ \bibnamefont
  {Ratner}},\ }\href@noop {} {\bibfield  {journal} {\bibinfo  {journal} {J.
  Chem. Phys.}\ }\textbf {\bibinfo {volume} {124}},\ \bibinfo {pages} {044108}
  (\bibinfo {year} {2006})}\BibitemShut {NoStop}%
\bibitem [{\citenamefont {Huo}\ and\ \citenamefont
  {Coker}(2012)}]{doi:10.1080/00268976.2012.684896}%
  \BibitemOpen
  \bibfield  {author} {\bibinfo {author} {\bibfnamefont {P.}~\bibnamefont
  {Huo}}\ and\ \bibinfo {author} {\bibfnamefont {D.~F.}\ \bibnamefont
  {Coker}},\ }\href@noop {} {\bibfield  {journal} {\bibinfo  {journal} {Mol.
  Phys.}\ }\textbf {\bibinfo {volume} {110}},\ \bibinfo {pages} {1035}
  (\bibinfo {year} {2012})}\BibitemShut {NoStop}%
\bibitem [{\citenamefont {Duke}\ and\ \citenamefont
  {Ananth}(2015)}]{doi:10.1021/acs.jpclett.5b01957}%
  \BibitemOpen
  \bibfield  {author} {\bibinfo {author} {\bibfnamefont {J.~R.}\ \bibnamefont
  {Duke}}\ and\ \bibinfo {author} {\bibfnamefont {N.}~\bibnamefont {Ananth}},\
  }\href@noop {} {\bibfield  {journal} {\bibinfo  {journal} {J. Phys. Chem.
  Lett.}\ }\textbf {\bibinfo {volume} {6}},\ \bibinfo {pages} {4219} (\bibinfo
  {year} {2015})}\BibitemShut {NoStop}%
\bibitem [{\citenamefont {Longuet-Higgins}(1975)}]{Longuet-Higgins147}%
  \BibitemOpen
  \bibfield  {author} {\bibinfo {author} {\bibfnamefont {H.~C.}\ \bibnamefont
  {Longuet-Higgins}},\ }\href@noop {} {\bibfield  {journal} {\bibinfo
  {journal} {P. Roy. Soc. Lond. A Mat.}\ }\textbf {\bibinfo {volume} {344}},\
  \bibinfo {pages} {147} (\bibinfo {year} {1975})}\BibitemShut {NoStop}%
\bibitem [{\citenamefont {Berry}(1984)}]{Berry45}%
  \BibitemOpen
  \bibfield  {author} {\bibinfo {author} {\bibfnamefont {M.~V.}\ \bibnamefont
  {Berry}},\ }\href@noop {} {\bibfield  {journal} {\bibinfo  {journal} {P. Roy.
  Soc. Lond. A Mat.}\ }\textbf {\bibinfo {volume} {392}},\ \bibinfo {pages}
  {45} (\bibinfo {year} {1984})}\BibitemShut {NoStop}%
\bibitem [{\citenamefont {Webster}, \citenamefont {Rossky},\ and\ \citenamefont
  {Friesner}(1991)}]{Webster1991494}%
  \BibitemOpen
  \bibfield  {author} {\bibinfo {author} {\bibfnamefont {F.}~\bibnamefont
  {Webster}}, \bibinfo {author} {\bibfnamefont {P.}~\bibnamefont {Rossky}}, \
  and\ \bibinfo {author} {\bibfnamefont {R.}~\bibnamefont {Friesner}},\
  }\href@noop {} {\bibfield  {journal} {\bibinfo  {journal} {Comput. Phys.
  Commun.}\ }\textbf {\bibinfo {volume} {63}},\ \bibinfo {pages} {494 }
  (\bibinfo {year} {1991})}\BibitemShut {NoStop}%
\bibitem [{\citenamefont {Hammes-Schiffer}\ and\ \citenamefont
  {Tully}(1994)}]{hammes-schiffer:4657}%
  \BibitemOpen
  \bibfield  {author} {\bibinfo {author} {\bibfnamefont {S.}~\bibnamefont
  {Hammes-Schiffer}}\ and\ \bibinfo {author} {\bibfnamefont {J.~C.}\
  \bibnamefont {Tully}},\ }\href@noop {} {\bibfield  {journal} {\bibinfo
  {journal} {J. Chem. Phys.}\ }\textbf {\bibinfo {volume} {101}},\ \bibinfo
  {pages} {4657} (\bibinfo {year} {1994})}\BibitemShut {NoStop}%
\bibitem [{\citenamefont {Prezhdo}\ and\ \citenamefont
  {Rossky}(1997)}]{Prezhdo:1997}%
  \BibitemOpen
  \bibfield  {author} {\bibinfo {author} {\bibfnamefont {O.~V.}\ \bibnamefont
  {Prezhdo}}\ and\ \bibinfo {author} {\bibfnamefont {P.~J.}\ \bibnamefont
  {Rossky}},\ }\href@noop {} {\bibfield  {journal} {\bibinfo  {journal} {J.
  Chem. Phys.}\ }\textbf {\bibinfo {volume} {107}},\ \bibinfo {pages} {825}
  (\bibinfo {year} {1997})}\BibitemShut {NoStop}%
\bibitem [{\citenamefont {Stock}\ and\ \citenamefont
  {Thoss}(1997)}]{PhysRevLett.78.578}%
  \BibitemOpen
  \bibfield  {author} {\bibinfo {author} {\bibfnamefont {G.}~\bibnamefont
  {Stock}}\ and\ \bibinfo {author} {\bibfnamefont {M.}~\bibnamefont {Thoss}},\
  }\href@noop {} {\bibfield  {journal} {\bibinfo  {journal} {Phys. Rev. Lett.}\
  }\textbf {\bibinfo {volume} {78}},\ \bibinfo {pages} {578} (\bibinfo {year}
  {1997})}\BibitemShut {NoStop}%
\bibitem [{\citenamefont {Sun}\ and\ \citenamefont {Miller}(1997)}]{sun:6346}%
  \BibitemOpen
  \bibfield  {author} {\bibinfo {author} {\bibfnamefont {X.}~\bibnamefont
  {Sun}}\ and\ \bibinfo {author} {\bibfnamefont {W.~H.}\ \bibnamefont
  {Miller}},\ }\href@noop {} {\bibfield  {journal} {\bibinfo  {journal} {J.
  Chem. Phys.}\ }\textbf {\bibinfo {volume} {106}},\ \bibinfo {pages} {6346}
  (\bibinfo {year} {1997})}\BibitemShut {NoStop}%
\bibitem [{\citenamefont {Wu}\ and\ \citenamefont {Herman}(2005)}]{Wu:2005kx}%
  \BibitemOpen
  \bibfield  {author} {\bibinfo {author} {\bibfnamefont {Y.}~\bibnamefont
  {Wu}}\ and\ \bibinfo {author} {\bibfnamefont {M.~F.}\ \bibnamefont
  {Herman}},\ }\href@noop {} {\bibfield  {journal} {\bibinfo  {journal} {J.
  Chem. Phys.}\ }\textbf {\bibinfo {volume} {123}},\ \bibinfo {pages} {144106}
  (\bibinfo {year} {2005})}\BibitemShut {NoStop}%
\bibitem [{\citenamefont {Miller}(2009)}]{doi:10.1021/jp809907p}%
  \BibitemOpen
  \bibfield  {author} {\bibinfo {author} {\bibfnamefont {W.~H.}\ \bibnamefont
  {Miller}},\ }\href@noop {} {\bibfield  {journal} {\bibinfo  {journal} {J.
  Phys. Chem. A}\ }\textbf {\bibinfo {volume} {113}},\ \bibinfo {pages} {1405}
  (\bibinfo {year} {2009})}\BibitemShut {NoStop}%
\bibitem [{\citenamefont {Huo}\ and\ \citenamefont
  {Coker}(2011)}]{:/content/aip/journal/jcp/135/20/10.1063/1.3664763}%
  \BibitemOpen
  \bibfield  {author} {\bibinfo {author} {\bibfnamefont {P.}~\bibnamefont
  {Huo}}\ and\ \bibinfo {author} {\bibfnamefont {D.~F.}\ \bibnamefont
  {Coker}},\ }\href@noop {} {\bibfield  {journal} {\bibinfo  {journal} {J.
  Chem. Phys.}\ }\textbf {\bibinfo {volume} {135}},\ \bibinfo {pages} {201101}
  (\bibinfo {year} {2011})}\BibitemShut {NoStop}%
\bibitem [{\citenamefont {Cotton}, \citenamefont {Igumenshchev},\ and\
  \citenamefont
  {Miller}(2014)}]{:/content/aip/journal/jcp/141/8/10.1063/1.4893345}%
  \BibitemOpen
  \bibfield  {author} {\bibinfo {author} {\bibfnamefont {S.~J.}\ \bibnamefont
  {Cotton}}, \bibinfo {author} {\bibfnamefont {K.}~\bibnamefont
  {Igumenshchev}}, \ and\ \bibinfo {author} {\bibfnamefont {W.~H.}\
  \bibnamefont {Miller}},\ }\href@noop {} {\bibfield  {journal} {\bibinfo
  {journal} {J. Chem. Phys.}\ }\textbf {\bibinfo {volume} {141}},\ \bibinfo
  {pages} {084104} (\bibinfo {year} {2014})}\BibitemShut {NoStop}%
\bibitem [{\citenamefont {Lopreore}\ and\ \citenamefont
  {Wyatt}(1999)}]{PhysRevLett.82.5190}%
  \BibitemOpen
  \bibfield  {author} {\bibinfo {author} {\bibfnamefont {C.~L.}\ \bibnamefont
  {Lopreore}}\ and\ \bibinfo {author} {\bibfnamefont {R.~E.}\ \bibnamefont
  {Wyatt}},\ }\href@noop {} {\bibfield  {journal} {\bibinfo  {journal} {Phys.
  Rev. Lett.}\ }\textbf {\bibinfo {volume} {82}},\ \bibinfo {pages} {5190}
  (\bibinfo {year} {1999})}\BibitemShut {NoStop}%
\bibitem [{\citenamefont {Hu}\ \emph {et~al.}(2000)\citenamefont {Hu},
  \citenamefont {Ho}, \citenamefont {Rabitz},\ and\ \citenamefont
  {Askar}}]{PhysRevE.61.5967}%
  \BibitemOpen
  \bibfield  {author} {\bibinfo {author} {\bibfnamefont {X.-G.}\ \bibnamefont
  {Hu}}, \bibinfo {author} {\bibfnamefont {T.-S.}\ \bibnamefont {Ho}}, \bibinfo
  {author} {\bibfnamefont {H.}~\bibnamefont {Rabitz}}, \ and\ \bibinfo {author}
  {\bibfnamefont {A.}~\bibnamefont {Askar}},\ }\href@noop {} {\bibfield
  {journal} {\bibinfo  {journal} {Phys. Rev. E}\ }\textbf {\bibinfo {volume}
  {61}},\ \bibinfo {pages} {5967} (\bibinfo {year} {2000})}\BibitemShut
  {NoStop}%
\bibitem [{\citenamefont {Kendrick}(2003)}]{kendrick:5805}%
  \BibitemOpen
  \bibfield  {author} {\bibinfo {author} {\bibfnamefont {B.~K.}\ \bibnamefont
  {Kendrick}},\ }\href@noop {} {\bibfield  {journal} {\bibinfo  {journal} {J.
  Chem. Phys.}\ }\textbf {\bibinfo {volume} {119}},\ \bibinfo {pages} {5805}
  (\bibinfo {year} {2003})}\BibitemShut {NoStop}%
\bibitem [{\citenamefont {Wyatt}\ and\ \citenamefont
  {Trahan}(2006)}]{wyatt2006quantum}%
  \BibitemOpen
  \bibfield  {author} {\bibinfo {author} {\bibfnamefont {R.}~\bibnamefont
  {Wyatt}}\ and\ \bibinfo {author} {\bibfnamefont {C.}~\bibnamefont {Trahan}},\
  }\href@noop {} {\emph {\bibinfo {title} {Quantum Dynamics with Trajectories:
  Introduction to Quantum Hydrodynamics}}},\ Interdisciplinary applied
  mathematics\ (\bibinfo  {publisher} {Springer},\ \bibinfo {year}
  {2006})\BibitemShut {NoStop}%
\end{thebibliography}%

\end{document}